\documentclass[aps,prd,preprint,tightening,showpacs]{revtex4-2}
\usepackage{amsmath,amssymb,amsthm,graphicx}
\usepackage[mathscr]{eucal}
\usepackage[colorlinks,linkcolor=blue,
anchorcolor=blue,citecolor=green]{hyperref}
\usepackage[dvipsnames,usenames]{color}

\newcommand{\be}{\begin{equation}}
\newcommand{\ee}{\end{equation}}
\newcommand{\ba}{\begin{eqnarray}}
\newcommand{\ea}{\end{eqnarray}}
\usepackage{epsf,epsfig,graphics}
\usepackage{verbatim,color,ulem}
\begin{document}
\title{Homoclinic orbits in  Kerr-Newman black holes}
\author{Yi-Ting Li}
\author{Chen-Yu Wang}
\author{Da-Shin Lee}
\email{dslee@gms.ndhu.edu.tw}
\author{Chi-Yong Lin}
\email{lcyong@gms.ndhu.edu.tw}
\affiliation{
Department of Physics, National Dong Hwa University, Hualien, Taiwan, Republic of China}
\date{\today}

\begin{abstract}
We present the exact solutions of the homoclinic orbits for the timelike geodesics of the particle on the general nonequatorial orbits in the Kerr-Newman black holes.
The homoclinic orbit is the separatrix between bound and plunging geodesics, a solution that asymptotes to an energetically bound, unstable spherical orbit.
The solutions are written in terms of  the elliptical integrals and the Jacobi elliptic functions of manifestly real functions of the Mino time where we focus on the effect from the charge of the black hole to the homoclinic orbits.
The parameter space of the homoclinic solutions is explored.
{The nonequatorial homoclinic orbits in Kerr cases can be obtained by setting the charge of the black holes to be zero.}
The homoclinic orbits and the associated phase portrait as a function of the radial position and its derivation with respect to the Mino time are plotted using the analytical solutions.
In particular, the solutions can reduce to the zero azimuthal angular moment homoclinic orbits for understanding the frame dragging effects from the spin as well as the charge of the black hole.
The implications of the obtained results to observations are discussed.

\end{abstract}


\maketitle

\newpage

\section{Introduction}
Recently successive observations of gravitational waves emitted by the merging of binary systems provide one of the long-awaited confirmations of General Relativity (GR) \cite{LIGOS:2016,LIGOS:2018,LIGOS:2020}.
The capture of the spectacular images of the supermassive black holes M87* at the center of the M87 galaxy \cite{M87:2019_1}
and Sgr A* at the center of our galaxy \cite{SgrA:2022_1}
is also a great achievement that provides direct evidence of the existence of the black holes, the solutions of Einstein's field equations \cite{MIS,CHAS}.
Spiraling of stellar mass compact objects into supermassive black holes due to the backreaction of gravitational wave (GW) emission is predicted to be a key source of low frequency gravitational waves, which will be targeted for the planned space-based Laser Interferometer Space Antenna (LISA) \cite{eLISA,Bar}.
Some of the science goals toward  the direct detection of these extreme mass ratio inspirals (EMRIs) are to precisely determine the properties of the EMRIs black hole and its inspiraling components to extract the black hole's astrophysical environments \cite{Bab,Koc}.

The zeroth order approximation in the {extreme} mass ratio is  considered as the small body travels along the geodesic of the background spacetime of the massive black hole.
The beginning of the extensive study of the timelike geodesics near the black holes dates back to a remarkable discovery from Carter of the so-called Carter constant \cite{Carter}.
In the family of the Kerr black holes
the geodesics of the particle due to the spacetime symmetry of the Kerr family possesses two conserved quantities, the energy {$E_{m}$} and the azimuthal angular momentum {$L_{m}$} of the particle.
Nevertheless, the existence of the third conserved quantity Carter constant renders the geodesic equations as a set of first-order differential equations.
%
Later, the introduction of the Mino time \cite{Mino_2003} further fully decouples the geodesics equations with the solutions expressed in terms of the elliptical functions.
In this work,  we would like to particularly focus on the geodesic dynamics in the case of Kerr-Newman black holes.
The Kerr-Newman metric of the solution of the Einstein-Maxwell equations represents a generalization of the Kerr metric, and describes spacetime in the exterior of a rotating charged black hole where, in addition to gravitation fields, both electric and magnetic fields exist intrinsically from the black holes.
Although one might not expect that astrophysical black holes have a large residue electric charge, some accretion scenarios were proposed to investigate the possibility of spinning charged back holes \cite{Dam_1978}.
Moreover, theoretical considerations, together with recent observations of structures near Sgr A* by the GRAVITY experiment \cite{Abu_2018}, indicate the possible presence of a small electric charge of central supermassive black hole \cite{Zaj_2018,Zaj_2019}.
Thus, it is still of great interest to explore the geodesic dynamics in the Kerr-Newman black hole.

In our previous paper \cite{Wang_2022}, we have studied the null and timelike geodesics of the light and the neutral particles respectively in the exterior of Kerr-Newman  black holes. We classify the roots for both angular and radial potentials, and mainly focus on those of the radial potential with an emphasis on the effect from the charge of the black holes.
We then obtain the solutions of the trajectories in terms of  the elliptical integrals and the Jacobi elliptic functions for the geodesics, which are manifestly real functions of the Mino time that the initial conditions can be explicitly specified.
In this paper, we will mainly focus on the work of \cite{Levin_2009}, which has showed the exact Homoclinic  solutions on the equatorial plane in the Kerr black holes, and extend the solutions to a general nonequatorial orbit in the Kerr-Newman black hole with an additional charge.
%
{By taking the zero charge limit, the obtained solutions can reduce to those of  the nonequatorial homonclinic orbits in the Kerr black holes of more relevance to astrophysical black holes.
%
 A homoclinic orbit is the separatrix between bound and plunging geodesics,  and is an orbit that asymptotes to an energetically bound, unstable spherical orbit. There are several approaches to model the EMRIs through a series of Kerr geodesics \cite{Gla,Dra1,Dra2,Dra3,Lang,Col,Dra4,Levin_2,Levin_3,Levin_4}.
  Due to the dissipation effects from the emission of gravitational radiation on EMRIs, the trajectories of the stellar mass compact object  will transit from
an inspiral to a plunge through a homoclinic orbit.
The exact solutions for the homoclinic orbits
 could be very useful for analytic or numerical studies of the transition from inspiral to plunge \cite{Sha,Spe}. A thorough knowledge of the underlying dynamics of EMRIs becomes essential to shape
the gravitational waveform emitted by them.}
Also, the added noise provided by the random kicks on the homoclinic orbit will expectedly give the chaotic behavior on the particle moving near the horizon \cite{Bom}.
On top of the fantastic observational effects, it is conjectured that there exists the bound of the Lyapunov exponent based upon the AdS/CFT correspondence \cite{Mald}.
The finding of the exact solution of the homoclinic orbits in the Kerr and Kerr-Newman black holes will be the first step to study the potential chaotic  motion when the particle travels near the horizon by explicitly computing the  Lyapunov exponents to justify or falsify the conjecture.
In this paper, we will start from the analytical solutions of the bound motion derived in \cite{Wang_2022}, and find the solutions of the homoclinic orbits where the particle starts off from the position of the largest root of the radial potential and spent tremendous time moving toward the position of the double root of the radial potential.
The parameter space of the homoclinic solutions will also be explored.

In Sec. \ref{secII}, we give a brief review of the timelike geodesic equations from which to define the radial and the angular potentials in terms of the Carter constant $C_m$ and the azimuthal angular momentum $L_m$ of the particle normalized by the energy $E_m$.
The parameter space to have the general nonequatorial homoclinic solutions is explored.  Later the results of the homoclinic orbits  are adapted from Appendixes which show the solutions of the $\theta$ and $r$ dependence in general bound motion.
We then particularly consider the  zero azimuthal angular momentum homoclinic orbits, discussing the frame dragging effects in {Sec. \ref{secIII}}.
{In {Sec. \ref{secIV}} the phase difference between the nonequatorial homoclinic orbit and the spherical motion of the unstable orbit is considered.}
In {Sec. \ref{secV}} the obtained solutions will reduce to the equatorial orbits to compare with \cite{Levin_2009}  in the Kerr black holes as the charge of the black hole in our case is set to zero.
All results will be summarized in the closing section.

\section{Equations of motion for timelike geodesics and Homoclinic orbits}\label{secII}

\subsection{Equations of motion}

We start by reviewing the dynamical equations of the particle in the Kerr-Newman black hole.
In the Boyer-Lindquist coordinates the metric for a Kerr-Newman black hole with the gravitational mass $M$ and spin parameter $a=J/M$ reads as \cite{MIS}
\begin{align}\label{ds_square}
ds^2=-\frac{\Delta}{\Sigma}\left(dt-a\sin^2\theta d\phi \right)^2+\frac{\sin^2\theta}{\Sigma}\left[(r^2+a^2)d\phi-a\,dt \right]^2+\frac{\Sigma}{\Delta}dr^2+\Sigma d\theta^2\;,
\end{align}
where
$\Sigma=r^2+a^2\cos^2\theta$ and $\Delta=r^2-2Mr+a^2+Q^2$.
(In this paper we use the geometrized units $G=c=1$.)
The outer/inner event horizons $r_{+}/r_{-}$ are obtained from $\Delta(r)=0$, giving
\be
r_{\pm}=M\pm\sqrt{M^2-({a^2+Q^2})}\;,
\ee
which requires $M^2 \ge {a^2+Q^2}$.

For the asymptotically flat, stationary and axial-symmetric black holes as in those of (\ref{ds_square}), the metric is independent of $t$ and $\phi$. The associated  Killing vectors can be written as
 \begin{align}
 \xi_{(t)}^\mu=\delta_t^\mu , \quad \xi_{\phi}^\mu=\delta_\phi^\mu \,.
 \end{align}
The  conserved quantities according to the above symmetry, namely the energy $E_m$ and the azimuthal angular momentum $L_m$ along a geodesic, can be constructed by the above Killing
vectors and the four velocity $u^\mu= dx^\mu /d\sigma_m $ defined in terms of the proper time $\sigma_m$,
 \begin{align}
 E_m & \equiv -\xi_{(t)}^\mu u_\mu ,\label{E_m}\\
L_m & \equiv \xi_{(\phi)}^\mu u_\mu  \label{L_m}\,.
 \end{align}
The third conservative quantity is the Carter constant written as follows
\begin{equation}\label{mathbb_C}
 C_m= \Sigma^2\left(u^{\theta}\right)^2-a^2E_m^2\cos^2\theta +L_m^2\cot^2\theta+a^2m^2\cos^{2}\theta\, .
 \end{equation}

Together with the  timelike geodesics of the particle,
 {$u^\mu u_\mu=-m^2$}, the equations of motion for the Boyer-Lindquist coordinates read as
\begin{align}
&{\Sigma\frac{d{r}}{d\sigma_m}}=\pm_r\sqrt{R_m({r})}\,, \label{r_eq_particle}\\
&{\Sigma\frac{d\theta}{d\sigma_m}}=\pm_{\theta}\sqrt{\Theta_m(\theta)}\,,\label{theta_eq_particle}\\
&{\Sigma\frac{d\phi}{d\sigma_m}}=\frac{{a}}{{\Delta}}\left[\left({r}^2+{a}^2\right)\gamma_m-{a}\lambda_m\right]-\frac{1}{\sin^{2}\theta}\left({a}\gamma_m \sin^2\theta-\lambda_m\right)\,, \label{phi_eq_particle}\\
&{\Sigma\frac{d{t}}{d\sigma_m}}=\frac{{r}^2+{a}^2}{{\Delta}}\left[\left({r}^2+{a}^2\right)\gamma_m-{a}\lambda_m\right]-{a}\left({a}\gamma_m \sin^2\theta-\lambda_m\right) \,,\label{t_eq_particle}
\end{align}
where
\begin{align}
&\gamma_m=\frac{E_m}{m},\hspace*{2mm}\lambda_m\equiv\frac{L_m}{m},\hspace*{2mm}\eta_m\equiv\frac{ {C}_m}{m^2}.
\end{align}
The conserved quantities, the energy, the azimuthal angular momentum, and the the Carter constant from (\ref{r_eq_particle}), (\ref{theta_eq_particle}), (\ref{phi_eq_particle}), and (\ref{t_eq_particle}) can be determined from the initial conditions.
In (\ref{r_eq_particle}) and (\ref{theta_eq_particle}) the symbols $\pm_r={\rm sign}\left(u^{r}\right)$ and $\pm_{\theta}={\rm sign}\left(u^{\theta}\right)$ are defined by four velocity of the particle. Also in these two equations,  the radial and angular potentials  $R_m({r})$ and $\Theta_m(\theta)$ are respectively obtained as
\begin{align}
&\label{R_Potential} R_m({r})=\left[\left({r}^2+{a}^2\right)\gamma_m-{a}\lambda_m\right]^2-{\Delta}\left[ \eta_m+\left({a}\gamma_m-\lambda_m\right)^2+{r}^2\right]\, ,\\
&\label{ThetaPotential}
\Theta_m(\theta)=\eta_m+{a}^2\gamma_m^2\cos^2\theta-\lambda_m^2\cot^2\theta-{a}^2\cos^2\theta \, .
\end{align}

It is known that all equations can be fully decoupled in terms of the so-called Mino time $\tau_m$ defined as \cite{Mino_2003}
\be
{\frac{dx^{\mu}}{d\tau_m}\equiv\Sigma\frac{dx^{\mu}}{d\sigma_m}}\,\label{tau'}\, .
\ee
For the source point $x_{i}^{\mu}$ and observer point $x^{\mu}$, the integral forms of the equations now becomes \cite{Gralla_2020a}
\ba
\tau_m-\tau_{mi}&=&{I_r}={G_\theta}  , \label{r_theta} \\
{\phi}-{\phi_i}&=& {I_\phi}+{\lambda_{m}} {G_\phi}\, , \label{phi}\\
{t}-{t_i}&=& {I_t}+a^2{\gamma_{m}}{G_t} \, ,\label{t}
\ea
where
\ba
I_{ r}&\equiv&\int_{r_{i}}^{r}\frac{1}{\pm_r\sqrt{R_m(r)}}dr,\quad G_{ \theta}\equiv\int_{\theta_{i}}^{\theta}\frac{1}{\pm_{\theta}\sqrt{\Theta_m(\theta)}}d\theta \, ,\\
I_{ \phi}&\equiv&\int_{r_{i}}^{r}\frac{{a\left[\left(2Mr-Q^2\right)\gamma_{m}-a\lambda_{m}\right]}}{\pm_r\Delta\sqrt{R_m(r)}}dr,\quad G_{ \phi}\equiv\int_{\theta_{i}}^{\theta}\frac{{\csc^2\theta}}{\pm_{\theta}\sqrt{\Theta_m(\theta)}}d\theta \, ,\label{I_G_phi}\\
I_{t}&\equiv&\int_{r_{i}}^{r}\frac{{r^2\gamma_{m}\Delta+(2Mr-Q^2)\left[\left(r^2+a^2\right)\gamma_{m}-a{\lambda_{m}}\right]}}{\pm_r\Delta\sqrt{R_m(r)}}dr,\;\; G_{t}\equiv\int_{\theta_{i}}^{\theta}\frac{\cos^2\theta}{\pm_{\theta}\sqrt{\Theta_m(\theta)}}d\theta  .
\ea
In \cite{Wang_2022}, we have shown the solutions to both null and timelike geodesics. In this work, we will mainly focus on the homoclinic orbits that can be achieved from the alternative expressions of the solutions of the bound motion in \cite{Wang_2022}, which is presented in Appendixes.

\begin{figure}[h]
 \centering
 \includegraphics[scale=0.46]{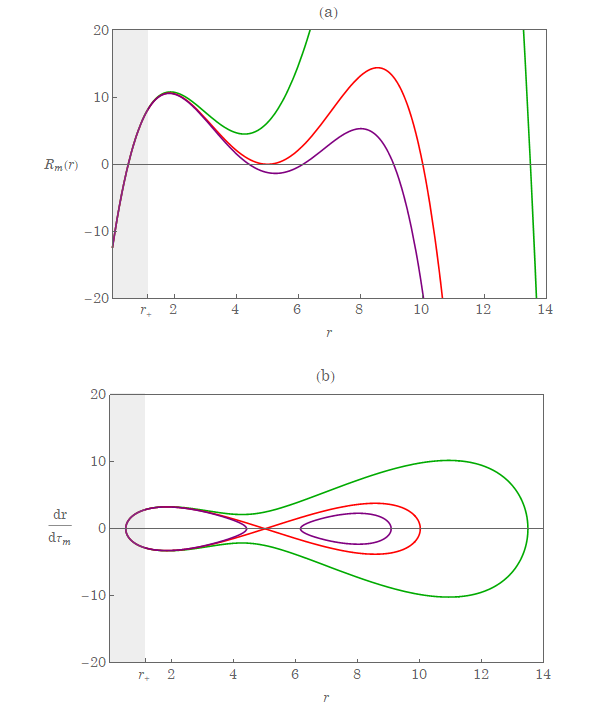}
 \caption{
 {
 \label{phase_portrait}
  The phase plane of the trajectories. The motions of plunging into the black hole {(green)}, the homoclinic solution (red) as well as the bound motion (purple) in the exterior of the Kerr-Newman black hole {($a/M=0.7, Q/M=0.7$)}.}}
 \end{figure}

{
Since the equations of motion are integrable, we thus can construct the diagrams of the exact trajectories on the equatorial plane in terms of $r$ and $dr/dt$, which reveal qualitative feature of the orbits \cite{Bom}.
%
{In Fig. \ref{phase_portrait}(a) the radial potential $R_m(r)$ in (\ref{R_Potential}) is plotted for a few representative parameters of the azimuthal angular momentum $\lambda_m$ and the Carter constant $\eta_m$ for the bound motion ($\gamma_m <1$).
The kinematically allowed regions for the particle to move are for $R_m(r) >0$.  Three types of motion can be seen from the plot.
The green curve is for the case that the particle starts from the largest root of $R_m(r)$, moves toward the black hole, and eventually plunges into it.
The purple curve is for the particle traveling around the black hole exterior.
However, the red curve lies in the separatrix between the two above mentioned motions known as the homoclinic orbit, where  the particle starts off from the position of the largest root of the radial potential and spends tremendous time moving toward the position of the double root of the radial potential.
The corresponding phase portrait is shown in Fig. \ref{phase_portrait}(b) given by the solutions of (\ref{r_eq_particle}). All the analytical solutions for such bound motion have been studied in \cite{Wang_2022}. Here we would like to focus on the analytical solutions of  nonequatorial homoclinic orbits and explore the parameter space {$\lambda_m$} and $\eta_m$ giving such orbits.   }

\subsection{Homoclinic orbits in the parameter space of $\lambda_m$ and $\eta_m$}\label{SecIIB}

{
The homoclinic orbits can be characterized by the parameters $\lambda_m$ and $\eta_m$ given by the double root of the radial potential, namely  $R_{m}\left(r\right)=R_{m}'\left(r\right)=0$ in (\ref{R_m}).  They are \cite{Wang_2022}
}
%
%
\begin{align}
&\lambda_{\rm mss}=\frac{\left[r_{\rm mss}\left(Mr_{\rm mss}-Q^2\right)-a^2M\right]\gamma_m-\Delta\left(r_{\rm mss}\right)\sqrt{r_{\rm mss}^2\left(\gamma_m^2-1\right)+Mr_{\rm mss}}}{a\left(r_{\rm mss}-M\right)}\;, \label{tilde_lambda_m}\\
&{\eta}_{\rm mss}=\frac{r_{\rm mss}}{a^2\left(r_{\rm mss}-M\right)^2}
\Big\{ r_{\rm mss}\left(Mr_{\rm mss}-Q^2\right)\left(a^2+Q^2-Mr_{\rm mss}\right)\gamma_m^2\Big.
\notag\\
&\quad\quad\quad\quad\quad\quad\quad+2\left(Mr_{\rm mss}-Q^2\right)\Delta\left(r_{\rm mss}\right)\gamma_m\sqrt{r_{\rm mss}^2\left(\gamma_m^2-1\right)+Mr_{\rm mss}}\notag\\
& \Big.
\quad\quad\quad\quad\quad\quad\quad\left.+\left[a^2\left(Mr_{\rm mss}-Q^2\right)-\left(\Delta\left(r_{\rm mss}\right)-a^2\right)^2\right]\left[r_{\rm mss}\left(\gamma_m^2-1\right)+M\right]\Big\} \right.\,, \label{tilde_eta_m}
\end{align}
where the subscript ``ss" means the spherical orbits with $s=\pm$, which indicates the two types of motion with respect to the relative sign between the black hole's spin and the azimuthal angular of the particle (see Sec. III C of \cite{Wang_2022}).

\begin{figure}[h]
 \centering
 \includegraphics[scale=0.6]{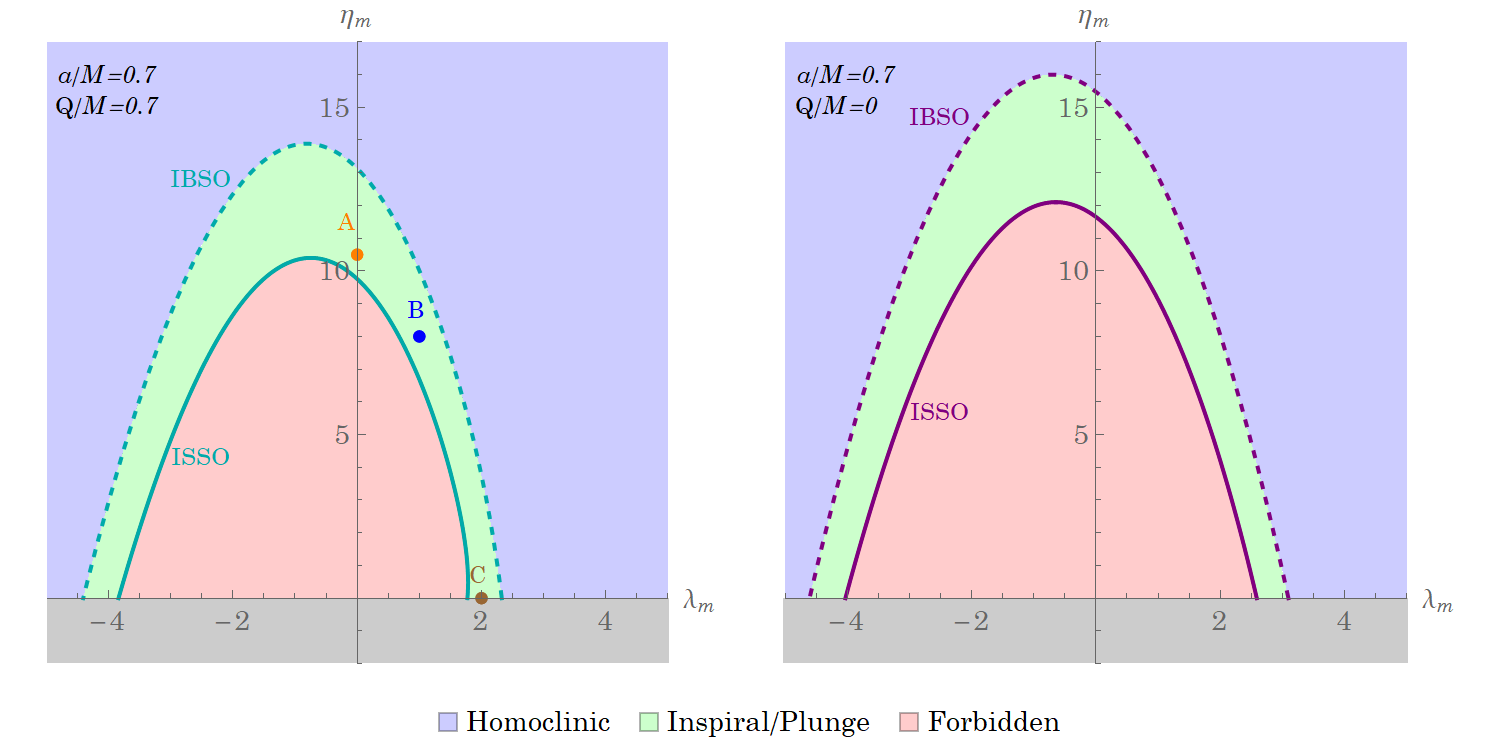}
 \caption{
 {
 \label{homoclinic_region_1}
 The plot shows curves of ISSO {(solid line)} and IBSO {(dashed line)} in the parameter space $\lambda_m$ and $\eta_m$, which demarcate the range of homoclinic solutions (light green). The particle with parameters in the light red and light blue regions plunges into the black hole or scatters by the black hole, respectively. The representative cases are as follows.
The parameters of case A can have  nonequatorial homoclinic solutions with the zero azimuthal angular momentum orbit where $\lambda_m=0$, case  B for the general nonzero $\lambda_m$ and $\eta_m$, and case C for equatorial homoclinic orbits with $\eta_m=0$ in \cite{Levin_2009}. {The corresponding curves are also shown for Kerr black holes with $Q=0$ for comparison.}
 }
}
\end{figure}

The spherical orbits with radius $r_{\rm mss}$  coincide with periastron, $r_{\rm mss}=r_p$, which eventually merges with the apastron $r_a$, the largest root of the radial potential, giving a triple root, also known as innermost stable spherical orbit (ISSO) by decreasing $\gamma_m$ \cite{Wang_2022}.
On the one hand, solving the triple root equations
\begin{equation}
    R_m(r)=0\,,\;\;R'_m(r)=0\,,\;\;R''_m(r)=0\,,
    \label{isso_eqs}
\end{equation}
one can demarcate the boundary of ISSO in the parameter space $\lambda_m$ and $\eta_m$.
On the other hand, the bound orbits are limited by the condition $\gamma_m \to 1$ or equivalently $r_a\to \infty$, known as innermost bound spherical orbit (IBSO).
With the double root conditions  together with $\gamma_m=1$,  one can find the boundary of IBSO.
The parameter region to have  homoclinic orbit is bound by the curves ISSO and IBCO in the parameter space $\lambda_m$ and $\eta_m$ as shown in the light green region in Fig. \ref{homoclinic_region_1}.
Obviously the particles with the parameters in the light blue (light red) region will evolve unbound (inspiral/plunge) motions.
Notice that case A is for the zero azimuthal angular momentum orbit.
%
The frame dragging effect due to the spin of the black hole drives the particle to rotate around the angle $\phi$, which will be discussed in Sec. \ref{secIII}.

The radii of $r_{\rm isso}$, the triple root of $R_m$ and $r_{\rm ibso}$, the radius of the double root together with $\gamma_m=1$ are given, respectively, by \cite{Wang_2022}
%
{
\begin{equation}\label{r_isso_noneq}
    -M r_{\rm isso}^5\Delta\left(r_{\rm isso}\right)+4\left(M r_{\rm isso}^3-Q^2r_{\rm isso}^2+a^2\eta_{\rm isso}-as\sqrt{\Gamma_{\rm Ms}}\right)^2=0
\end{equation}
%
and
{\begin{equation}\label{r_ibso_noneq}
-M r_{\rm ibso}^7+\left(2M r_{\rm ibco}^3 -Q^2 r_{\rm ibso}^2+a^2\eta_{\rm ibso}-a s\sqrt{\Gamma_{\rm Mb}}\right)^2=0\,,
\end{equation}}
\noindent{where}
\begin{align}
\Gamma_{\rm Ms/b}&=r_{\rm isso/ibso}^4\left(M r_{\rm isso/ibso}-Q^2\right)\notag\\
&-\eta_{\rm isso/ibso}\left[r_{\rm isso/ibso}\left(r_{\rm isso/ibso}-3M\right)+2Q^2\right]r_{\rm isso/ibso}^2+a^2\eta_{\rm isso/ibso}^2\geq0 \,. \label{|gamma_noneq}
\end{align}
The radius of the homoclinic orbit lies between the radii $r_{\rm isso}$ and $r_{\rm ibso}$ according to the values of $\lambda_m$ and $\eta_m$ of the parameter regions  in Fig. \ref{homoclinic_region_2}, in which we have shown the boundaries for various combinations of the black hole parameters $a$ and $Q$.

\begin{figure}[h]
 \centering
 \includegraphics[scale=0.4]{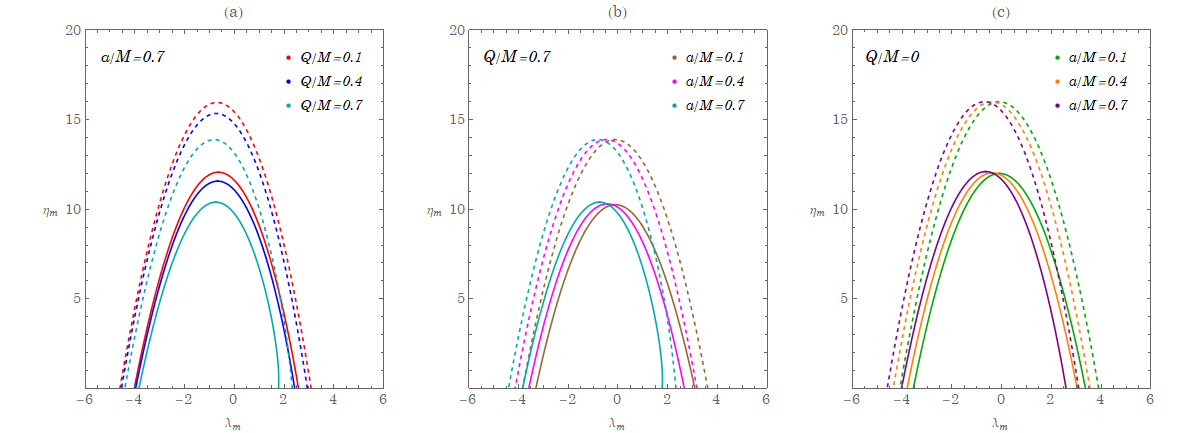}
 \caption{
 {
 \label{homoclinic_region_2}
{The curves ISSO (solid) and IBSO (dashed) in the $(\lambda_m,\eta_m)$ parameter space for various combinations of   $a/M$ and $Q/M$ for Kerr-Newman black holes as well as various choices of  $a/M$ for Kerr black holes with $Q=0$.}  }.}
 \end{figure}
Further detailed explanations for Figs. \ref{homoclinic_region_1} and \ref{homoclinic_region_2} have been discussed in  \cite{Wang_2022}.
In general, for a fixed value of $\eta_{m}$ and black hole spin $a$, since the finite charge of the black hole gives additional repulsive effects to prevent the particle from collapsing into the black hole, the  radii of both $r_{\rm isco}$ and $r_{\rm ibso}$ will decrease with the increase of the charge, resulting in smaller azimuthal angular momentum $\vert \lambda_m\vert$.
%
In addition, for fixed $\eta_m$,  the direct orbit will have smaller values of $\vert \lambda_m\vert$ as compared with the retrograde orbit. 
Based upon our results, for Sgr A* where the charge is more
likely below $10^{-18}$ \cite{Zaj_2018,Zaj_2019}, it is not expected to see noticeable difference from the  Kerr cases.

\subsection{Analytical solutions of homoclinic orbits in general nonequatorial cases}\label{sec_IIC}

One of the interesting trajectories with the parameters of the double root of the radial potential $R_m(r)$ is the homoclinic orbit. In this case, the particle starts from the point of the largest root $r_{m4}$, moves toward the black hole, and spends an infinity amount of time to reach the point of double roots $r_{m2}=r_{m3}=r_{\rm mss}$ with $\eta_{\rm mss}$ and $\lambda_{\rm mss}$ given by Eqs. (\ref{tilde_lambda_m}) and (\ref{tilde_eta_m}).
In this case, the analytical expressions summarized in Appendix \ref{app_B} gain tremendous simplifications since $r_{m2}=r_{m3}$ gives the parameter $k^B=1$, and the elliptical integrals and elliptic functions reduce to elementary functions \cite{Abramowitz}
%
\begin{align}
&F\left(\varphi|1\right)=\tanh^{-1}\left(\sin\varphi\right),\label{formula_1}\\
&E\left(\varphi|1\right)=\sin\varphi,\label{formula_2}\\
&\Pi\left(n;\varphi|1\right)=\frac{1}{n-1}\left[\sqrt{n}\tanh^{-1}\left(\sqrt{n}\sin\varphi\right)-\tanh^{-1}\left(\sin\varphi\right)\right],\label{formula_3}\\
&{\rm sn}\left(\varphi|1\right)=\tanh \varphi ,
\end{align}
where
$-\frac{\pi}{2}\leq\varphi\leq\frac{\pi}{2},\hspace*{4mm}-1<n<1$.

As for the coordinate $r(\tau_m)$, the homoclinic solution (\ref{r_tau_m_B}) and (\ref{X_B_tau_m}) simplify to the following formulas
%
\begin{equation}
r^{H}(\tau_m)=\frac{r_{m4}(r_{m1}-r_{\rm mss})-r_{m1}(r_{m4}-r_{\rm mss}){{\tanh}^2\left(X^{H}(\tau_m)\right)}}{(r_{m1}-r_{\rm mss})-(r_{m4}-r_{\rm mss}){\tanh}^2 X^{H}(\tau_m)}\label{r_tau_m_h}
\end{equation}
and
\begin{equation}
X^{H}(\tau_m)=\frac{\sqrt{\left(1-\gamma_m^2\right)({r_{\rm mss}}-r_{m1})(r_{m4}-{r_{\rm mss}})}}{2}\tau_m \, .
\end{equation}
The other relevant integrals resulting from the radial potentials that contribute to the solutions of the azimuthal angle $\phi(\tau_m)$ and the time $t(\tau_m)$ in (\ref{phi}) and (\ref{t}) are obtained from (\ref{I_phi_tau_m_b}) and (\ref{I_t_tau_m_b}) giving
 {\begin{equation}
I_{\phi}(\tau_m)=\frac{\gamma_m}{\sqrt{1-\gamma_m^2}}\frac{2Ma}{r_{+}-r_{-}}\left[\left(r_{+}-\frac{a}{2M}\frac{\lambda_m}{\gamma_m}-\frac{Q^2}{2M}\right)I_{+}^{H}(\tau_m)
-\left(r_{-}-\frac{a}{2M}\frac{\lambda_m}{\gamma_m}-\frac{Q^2}{2M}\right)I_{-}^{H}(\tau_m)\right]\quad\label{I_phi_tau_m_h}
\end{equation}
\begin{align}
&I_{t}(\tau_m)=\frac{\gamma_m}{\sqrt{1-\gamma_m^2}}\left\lbrace\frac{4M^2}{r_{+}-r_{-}}\left[\left(r_{+}-\frac{Q^2}{2M}\right)\left(r_{+}-\frac{a}{2M}\frac{\lambda_m}{\gamma_m}-\frac{Q^2}{2M}\right)I_{+}^{H}(\tau_m)\right.\right.\notag\\
&\quad\quad\left.\left.-\left(r_{-}-\frac{Q^2}{2M}\right)\left(r_{-}-\frac{a}{2M}\frac{\lambda_m}{\gamma_m}-\frac{Q^2}{2M}\right)I_{-}^{H}(\tau_m)\right]+2MI_{1}^{H}(\tau_m)+I_{2}^{H}(\tau_m) \label{I_t_tau_m_h}\right\rbrace+\left(4M^2-Q^2\right)\gamma_{m}\tau_m
\end{align}}
where
\begin{align}
&\label{I_pm_H}I_{\pm}^{H}(\tau_m)=\frac{2}{\sqrt{(r_{\rm mss}-r_{m1})(r_{m4}-r_{\rm mss})}}\left\lbrace\frac{1}{r_{m1}-r_{\pm}}X^{H}(\tau_m)\right.\notag\\
&\quad\quad\left.+\frac{r_{m1}-r_{m4}}{(r_{m1}-r_{\pm})\left(r_{m4}-r_{\pm}\right)\left(h_{\pm}-1\right)}\left[\sqrt{h_{\pm}}\tanh^{-1}\left(x^H\left(\tau_m\right)\sqrt{h_{\pm}}\right)-\tanh^{-1}\left(x^H\left(\tau_m\right)\right)\right]\right\rbrace\\
&I_{1}^{H}(\tau_m)=\frac{2}{\sqrt{(r_{\rm mss}-r_{m1})(r_{m4}-r_{\rm mss})}} \bigg\lbrace r_{m1}X^{H}(\tau_m)
\notag\\
&\quad\quad\quad\quad\quad\quad\quad\quad\quad\quad\quad\quad
+\frac{r_{m4}-r_{m1}}{h-1}\left[\sqrt{h}\tanh^{-1}\left(x^H\left(\tau_m\right)\sqrt{h}\right)-\tanh^{-1}\left(x^H\left(\tau_m\right)\right)\right] \bigg\rbrace
\label{I_1_H}
\end{align}
\begin{align}\label{I_2_H}
I_{2}^{{H}}(\tau_m)&=\frac{\sqrt{\left(r^{H}(\tau_m)-r_{m1}\right)\left(r^{H}(\tau_m)-r_{\rm mss}\right)^2\left(r_{m4}-r^{H}(\tau_m)\right)}}{r^{H}(\tau_m)-r_{m1}}\notag\\
&-\frac{r_{\rm mss}\left(r_{m4}-r_{m1}\right)-r_{m1}\left(r_{m4}+r_{m1}\right)}{\sqrt{(r_{\rm mss}-r_{m1})(r_{m4}-r_{\rm mss})}}X^{H}(\tau_m)\notag +\sqrt{(r_{\rm mss}-r_{m1})(r_{m4}-r_{\rm mss})}x^H (\tau_m) \notag\\
&+\frac{\left(r_{m4}-r_{m1}\right)\left(r_{m1}+2r_{\rm mss}+r_{m4}\right)}{\sqrt{(r_{\rm mss}-r_{m1})(r_{m4}-r_{\rm mss})}(h-1)}\left[\sqrt{h}\tanh^{-1}\left(x^H\left(\tau_m\right)\sqrt{h}\right)-\tanh^{-1}\left(x^H\left(\tau_m\right)\right)\right]
\end{align}
In the equations above we have introduced the notations
\begin{equation}\label{Upsilon_m_h}
x^H (\tau_m)=\sqrt{\frac{\left({r^{H}\left(\tau_m\right)}-r_{m4}\right)\left(r_{m1}-{r_{\rm mss}}\right)}{\left({r^{H}\left(\tau_m\right)}-r_{m1}\right)\left(r_{m4}-{r_{\rm mss}}\right)}}\;,
\end{equation}
\be\label{h_pm_h}
h_{\pm}=\frac{(r_{m1}-r_{\pm})(r_{m4}-r_{\rm mss})}{(r_{m4}-r_{\pm})(r_{m1}-r_{\rm mss})}\,,
\;\;h=\frac{r_{m4}-r_{\rm mss}}{r_{m1}-r_{\rm mss}} \, .
\ee
For a given Mino time $\tau_m$, the homoclinic orbits along the $r$ and $\theta$ directions can be drawn from the solutions of (\ref{theta_tau_m_a}) and (\ref{r_tau_m_h}).
Also, the corresponding $\phi$ and $t$ can be found through (\ref{phi}) and (\ref{t}) with the help of ${G_\phi}$ (\ref{G_phi_tau_m_a}), ${I_\phi}$ (\ref{I_phi_tau_m_h}), ${G_t}$ (\ref{G_t_tau_m_a}), and ${I_t}$ (\ref{I_t_tau_m_h}).
Together with the solutions along the $\theta$ direction, they are the homoclinic solutions for the general nonequatorial orbits  in the Kerr-Newman exterior, which are the main results of this work (see Fig. \ref{homoclinic_3d_case_B}). The above expressions can  further reduce  to  the Kerr black hole case by sending $Q\rightarrow 0$, and also on the equatorial plane \cite{Levin_2009}.

\begin{figure}[h]
 \centering
 \includegraphics[scale=0.4]{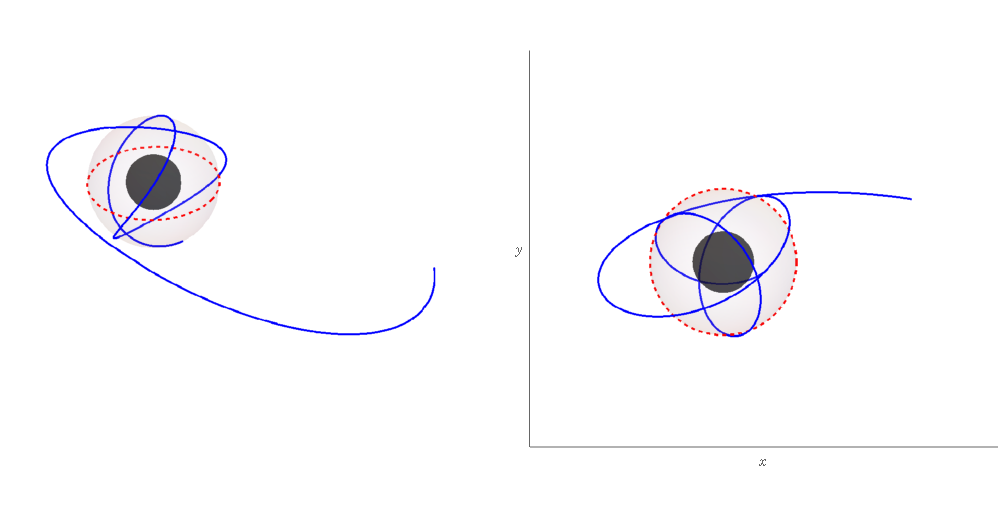}
 \caption{
 Illustration of a general nonequatorial homoclinic orbit.
 The particle with the parameters of case B given in Fig. \ref{homoclinic_region_1}  departs from $r_i=r_{m4}$, the largest root of the radial potential and approaches the unstable spherical orbit $r_{\rm mss}$ of the double root of the radial potential with characteristic angular frequencies shown in the text.
 { The red dotted circle is on the equatorial plane.}
 \label{homoclinic_3d_case_B}}
 \end{figure}

\section{zero azimuthal angular momentum homoclinic orbits}\label{secIII}

In Sec. \ref{secII} we have shown that the Kerr-Newman spacetimes permit homoclinic orbits in a limited region of parameter space $\lambda_m$ and $\eta_m$.
In particular, a particle with zero azimuthal angular momentum, $\lambda^A_m=0$, but $\eta_m^{\rm isso}<\eta^A_m <\eta_m^{\rm ibso}$, can realize homoclinic motion. Using the exact solutions discussed in Sec. II we show in Figs. \ref{homoclinic_zerolambda_90} and \ref{homoclinic_zerolambda_90_1} two illustrations for this type of motion.
The particles share the same set of parameters $\lambda_m$ and $\eta_m$, but with different  initial conditions.
With the zero azimuthal angular momentum, the motions show the change of $\phi$ simply due to the frame dragging effects from the black hole spin.
Additionally, the charge of the black hole will give another boost to the frame dragging to be seen later.

\begin{figure}[h]
 \centering
 \includegraphics[scale=0.45]{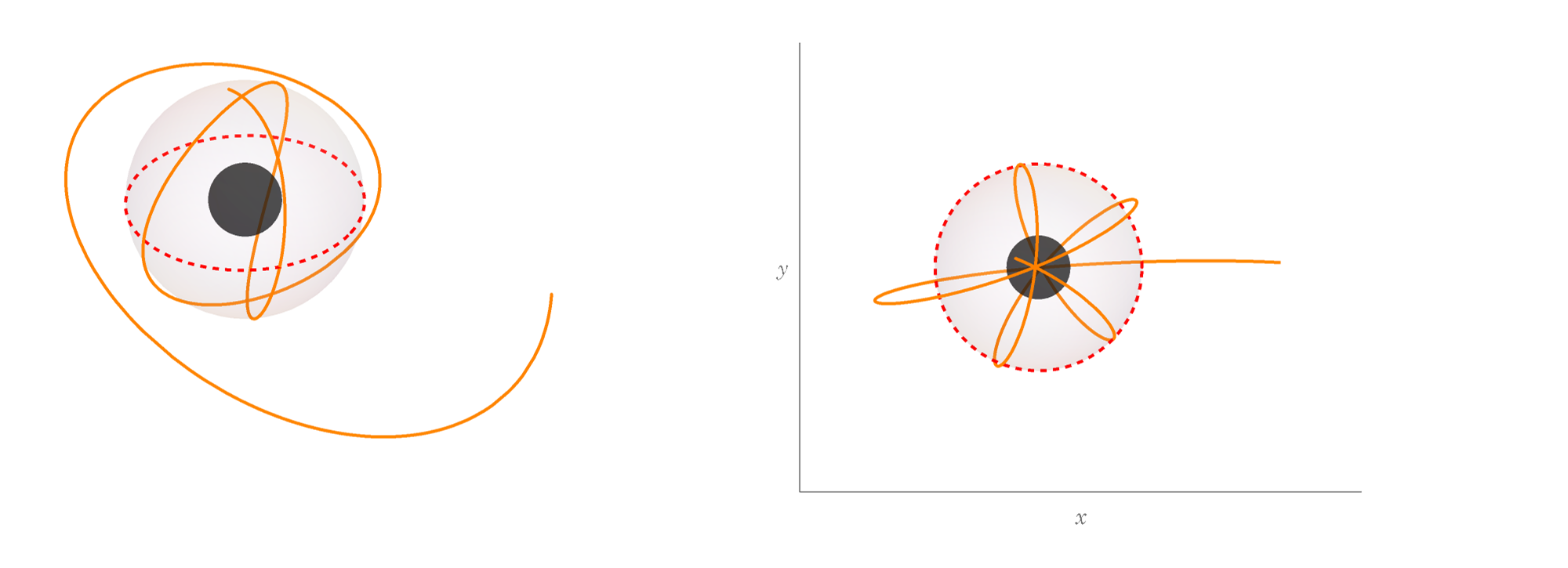}
 \caption{
 Illustration of a zero azimuthal angular momentum homoclinic orbit  with the parameters A in Fig. \ref{homoclinic_region_1} and  the initial conditions $r_i=r_{m4},\, \theta_i=\frac{\pi}{2}$. 
 { The red dotted circle is on the equatorial plane.}
 }
 \label{homoclinic_zerolambda_90}
 \end{figure}

The analytical solutions with zero azimuthal angular momentum have additional simplifications for the homoclinic orbits. The roots of the angular potential in (\ref{u_m})  reduce to
\begin{equation}
    u_{{m+}}=1\;, \quad u_{{m-}}=\frac{\eta_m}{a^2(1-\gamma^2_m)}\,.
    \label{upum}
\end{equation}
The relevant root is $u_{{m+}}=1$ giving the two turning points at the north and south poles of a sphere,  $\theta_{m+}=\cos^{-1}(-\sqrt{u_{m+}})={\pi}$ and $\theta_{m-}=\cos^{-1}(\sqrt{u_{m+}})={0}$, respectively \cite{Wang_2022}.
Substitute (\ref{upum}) into the solution of $\theta(\tau_m)$ in (\ref{theta_tau_m_a}) and notice that $u_{m+}/u_{m-}\ll 1$, since $\eta_m$ is relatively large value as seen in Fig.\ref{homoclinic_region_1}, also $a$ and $ 1-\gamma_m <1$. We can thus have the approximate solutions given by
\begin{align}\label{theta_tau_m_a_2}
\theta(\tau_m)&=\cos^{-1}\left(-\nu_{\theta_i}{\rm sn}
\left(\sqrt{\eta_m}
\left(\tau_m+\nu_{\theta_i}{\mathcal{G}_{\theta_i}}\right)\left|\frac{u_{m+}}
{u_{m-}}\right)\right.\right) \notag \\
&\approx\cos^{-1}\left(-\nu_{\theta_i}{\rm sin}\left(\sqrt{\eta_m}\left(\tau_m+\nu_{\theta_i}{\mathcal{G}_{\theta_i}}\right) \right)\right)\;.
\end{align}
 By inspection,  the  angular velocity in Mino time of the evolution around the $\theta$ can be read off as
\begin{equation}\label{omega_theta}
    \omega_\theta=\sqrt{\eta_m}\;.
\end{equation}
In Fig. \ref{homoclinic_zerolambda_90}, we choose  the initial conditions  $\theta_i=\frac{\pi}{2}$ and $\nu_{\theta_i}=1$, departing from the equator. Thus, with ${\mathcal{G}_{\theta_i}}=0$ in (\ref{g_theta_m_a}), the solution just becomes  $\theta(\tau_m)=\cos^{-1}\left(-{\rm sin}\left(\sqrt{\eta_m}\tau_m\right) \right)$.
And, in Fig \ref{homoclinic_zerolambda_90_1} the particle departs from the north pole with $\theta_i=0$ and $\nu_{\theta_i}=1$, giving ${\mathcal{G}_{\theta_i}}=\frac{\pi}{2\sqrt{\eta_m}}$. The evolution of coordinate $\theta$ follows $ \theta(\tau_m)=\cos^{-1}\left({\rm cos}\left(\sqrt{\eta_m}\tau_m\right) \right)$.

\begin{figure}[h]
 \centering
 \includegraphics[scale=0.45]{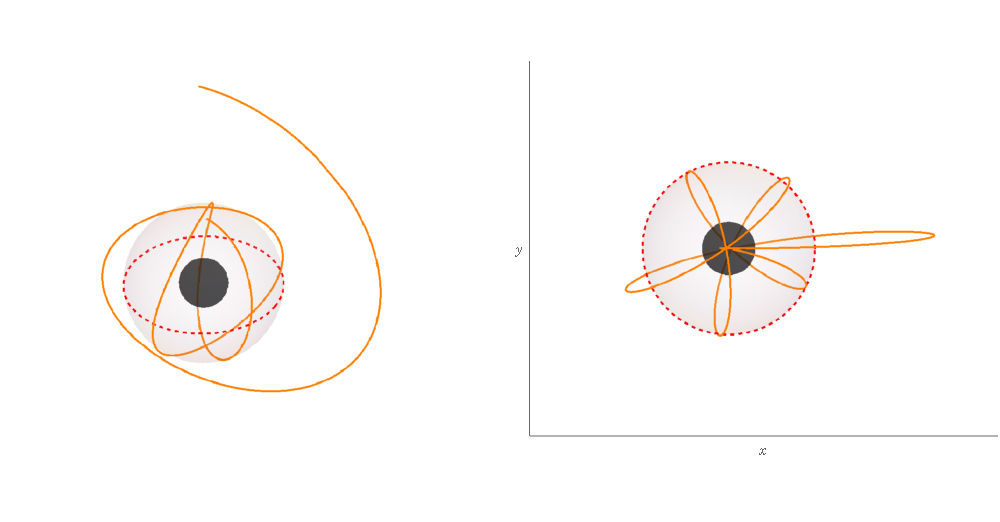}
 \caption{
 Illustration of a zero azimuthal angular momentum homoclinic orbit  with the parameters A in Fig. \ref{homoclinic_region_1} and  the initial conditions $r_i=r_{m4}, \theta_i=0$.
 {  The red dotted circle is on the equatorial plane.} } \label{homoclinic_zerolambda_90_1}
 \end{figure}

As for the azimuthal motion, we notice that the equation of motion in (\ref{phi_eq_particle}) reduces  to
\begin{equation}
   \frac{d\phi}{d\tau_m}= a\gamma_m \left[\frac{r^2+a^2}{\Delta(r)}-1\right]\,,
    \label{phi_eq_particle_zerolambda}
\end{equation}
simply due to the frame dragging.
In particular, the angular speed goes to a constant as $r\rightarrow r_{\rm mss}$ when $\tau_m \rightarrow \infty$ ($t\rightarrow \infty$), given by
\begin{equation} \label{omega_phi}
     a\gamma_m \left[\frac{r_{\rm mss}^2+a^2}{\Delta(r_{\rm mss})}-1\right]\equiv\omega_\phi \,,
     \quad
     \phi(\tau_m)=\omega_\phi \tau_m\,,
\end{equation}
which can also be obtained directly from the analytical formulas (\ref{phi}) and (\ref{I_phi_tau_m_h}) in the limits of $\lambda_m=0$ and $\tau_m \rightarrow \infty$ ($t\rightarrow \infty$).
This allows one to plot the ratio of $\omega_\phi$ to $\omega_\theta$ shown in Fig. \ref{homoclinic_omega_ratio}. It is expected that the larger value of the black hole spin $a$ gives relatively large dragging effects and moreover, the charge $Q$ of the black hole gives the boost to it. This is consistent with the conclusion in the study of the particle boomerang in \cite{Wang_2022}.
In there, we consider the spherical orbits of the particle with the radius of $r_{\rm mss}$ of the double root of the radial potential.
For sufficiently large value of black hole spin $a$ there can be a particle boomerang that leaves the north polar axis ($\theta = 0$) and travels along a constant-$r$ orbit to return to the north polar axis in precisely the opposite direction to that at which it leaves.
The dragging of inertial frames rotates the particle propagation direction by an angle $\Delta \phi= \pi$ during one orbit that returns to the same location on the north polar axis.
However, in the presence of the charge of the black hole in the Kerr-Newman black holes, the finite charge effect will enhance the frame dragging so as to reduce the needed value of the spin a to sustain $\Delta \phi= \pi$.

\begin{figure}[h]
 \centering
 \includegraphics[scale=0.4]{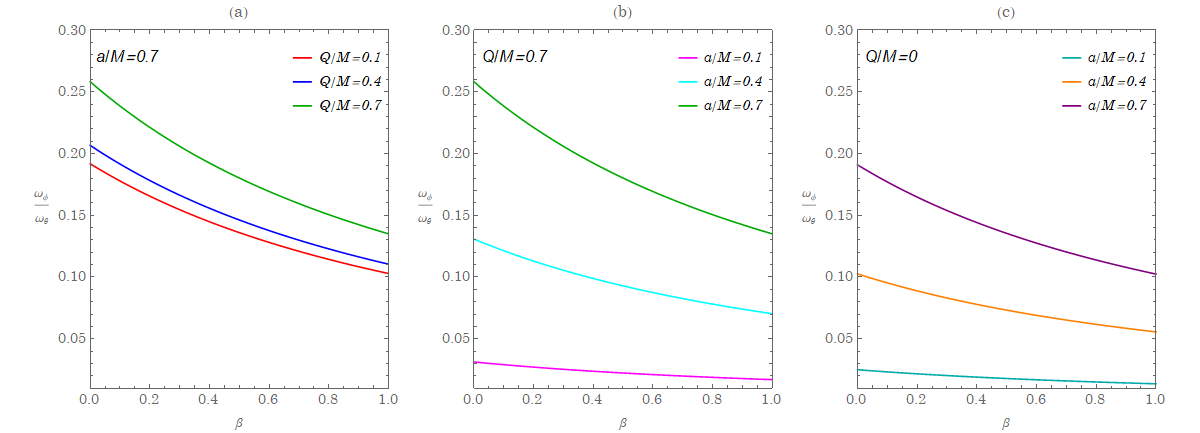}
 \caption{
 The ratio of $\omega_\phi$ to $\omega_\theta$ given in (\ref{omega_theta}) and (\ref{omega_phi}) respectively as a function of $\beta=\frac{r-r_{\rm isco}}{r_{\rm ibco}-r_{\rm isco}}$
 { for Kerr-Newman black holes as well as Kerr black holes.}   \label{homoclinic_omega_ratio}}
 \end{figure}

\section{{phase difference between nonequatorial Homoclinic orbits and spherical motion}}\label{secIV}}

According to \cite{Levin_2009}, when the homoclinic orbits on the equatorial plane in Kerr black holes are considered, it is found a nontrivial fact that although as $r$ approaches to the unstable orbit,
the accumulated phase $\phi$ and time spent go to infinity.
Nevertheless, the phase difference between the homoclinic orbits and the circular motion of the unstable orbits with the same radius of the double root turns out to be finite at time $t\rightarrow \infty$, when both start at the same initial time.
This will be also true for the homoclinic solution in the nonequatorial plane in the Kerr-Newman black holes where the circular motion is replaced by the spherical motion.
As $\tau_m \rightarrow \infty$, giving $t\rightarrow \infty$ and $r\rightarrow r_{\rm mass}$, the radius of the unstable orbit, with the associated values of $\lambda_{\rm mass},\eta_{\rm mass}$ obtained from (\ref{tilde_lambda_m}) and (\ref{tilde_eta_m}), the phase difference becomes
%
{\begin{align} \label{phase_diff_tau}
&\Delta \phi_{\tau_{m}}^{\rm diff}=\phi^{\rm sphe}-\phi_m \notag\\
&=\left\lbrace\frac{a\left[\left(2M r_{\rm mss}-Q^2\right)-a \lambda_{m}\right]}{\Delta(r_{\rm mss})}\tau_{m}-I_{\phi}(\tau_{m})\right\rbrace\Bigg|_{\tau_{m}\rightarrow\infty}\notag\\
&=\frac{2}{r_{+}-r_{-}}\frac{\left(2M a\gamma_{m}-r_{-}\lambda_{m}\right)r_{+}-Q^2\left(a\gamma_{m}-\lambda_{m}\right)}{\left(r_{{\rm mss}}-r_{+}\right)\sqrt{\left(1-\gamma_{m}^2\right)\left(r_{+}-r_{m1}\right)\left(r_{m4}-r_{+}\right)}}\tanh^{-1}\sqrt{\frac{\left(r_{+}-r_{m1}\right)\left(r_{m4}-r_{\rm mss}\right)}{\left(r_{m4}-r_{+}\right)\left(r_{\rm mss}-r_{m1}\right)}}\notag\\
&-\frac{2}{r_{+}-r_{-}}\frac{\left(2M a\gamma_{m}-r_{+}\lambda_{m}\right)r_{-}-Q^2\left(a\gamma_{m}-\lambda_{m}\right)}{\left(r_{{\rm mss}}-r_{-}\right)\sqrt{\left(1-\gamma_{m}^2\right)\left(r_{-}-r_{m1}\right)\left(r_{m4}-r_{-}\right)}}\tanh^{-1}\sqrt{\frac{\left(r_{-}-r_{m1}\right)\left(r_{m4}-r_{\rm mss}\right)}{\left(r_{m4}-r_{-}\right)\left(r_{\rm mss}-r_{m1}\right)}}\, .
\end{align}}
The general solutions of the homoclinic orbits and the spherical orbit contain the radial $r$ and angular $\theta$ dependence seen from Eqs.(\ref{phi}) and (\ref{I_G_phi}). The $\theta$ dependence cancels in computing $\Delta \phi_{\tau_{m}}^{\rm diff}$ and the only contributions come from the radial $r$ dependence.
To obtain (\ref{phase_diff_tau}), we have used the $r$ dependent solution of $\phi$  (\ref{I_phi_tau_m_h}) to the homoclinic orbits.
Also,  the radius of the spherical orbit is set at $r=r_{\rm mass}$ with the $r$ dependent solution  obtained from (\ref{phi_eq_particle})  in terms of the Mino time in (\ref{tau'}).
For $r_{\rm mss}$ lying between $r_{\rm isso}$ and $r_{\rm ibco}$ determined by (\ref{r_isso_noneq}) and (\ref{r_ibso_noneq}),
the corresponding $\lambda_{\rm isso}, \eta_{\rm isso}$ and $\lambda_{\rm ibso}, \eta_{\rm ibso}$ are determined by (\ref{tilde_lambda_m}) and (\ref{tilde_eta_m}).

%
\begin{figure}[h]
 \centering
\includegraphics[width=0.99\columnwidth=0.99,trim=20 400 20 400,clip]{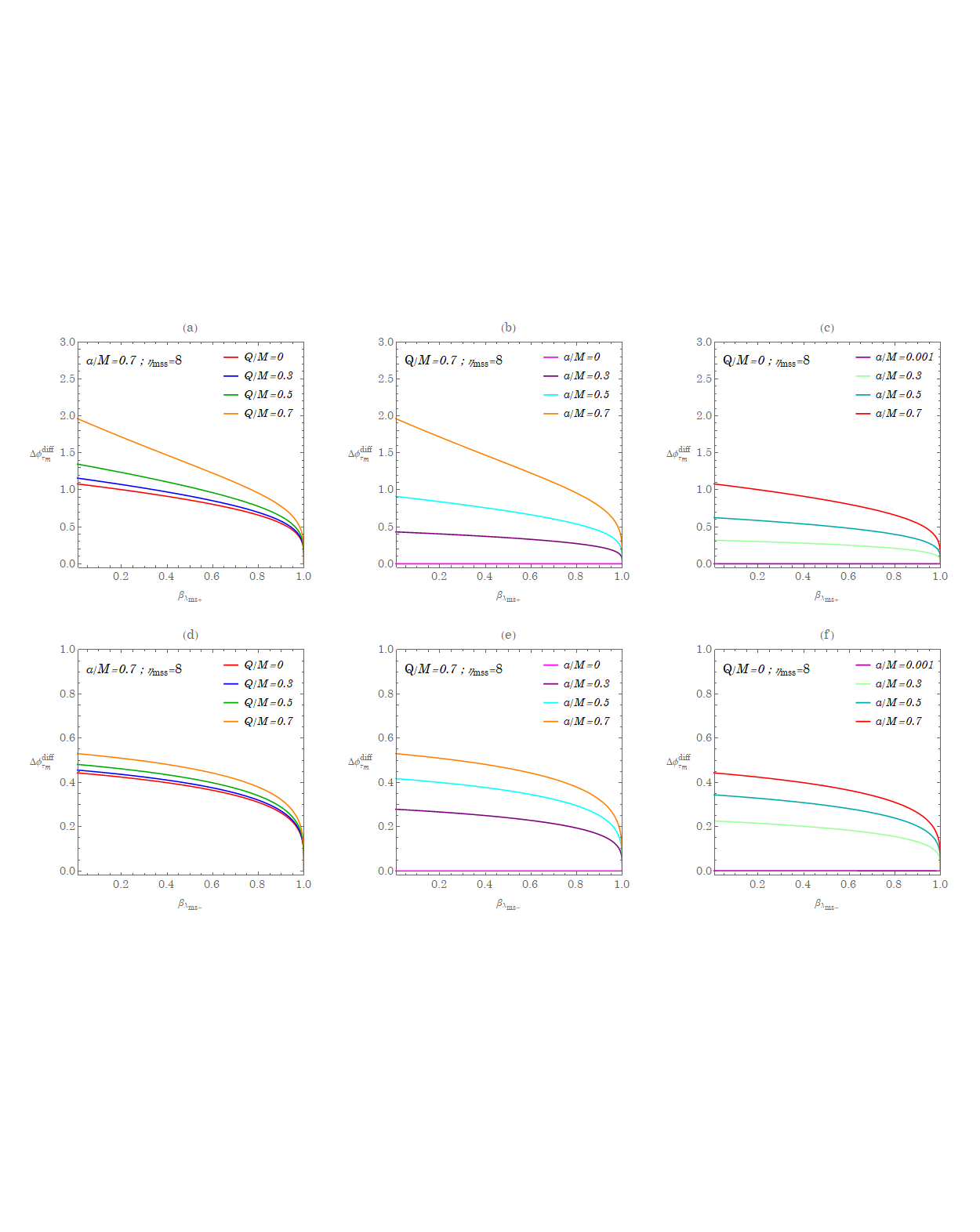}
 \caption{
{Phase difference $\Delta \phi_{\tau_{m}}^{\rm diff}=\phi^{\rm sphe}-\phi_m$ as a function of $\beta_{\lambda_{\rm mss}}=\frac{|\lambda_{\rm ibso}|-|\lambda_{\rm mss}|}{|\lambda_{\rm ibso}|-|\lambda_{\rm isso}|}$ in (\ref{phase_diff_tau}) for nonequatorial solutions.
The upper plots are for the direct motion and the lower plots are for the retrograde motion. }}\label{Phase_diff}
 \end{figure}

%
In Fig. \ref{Phase_diff}, the phase difference is plotted as a function of $\beta_{\lambda_{\rm mass}}$ defined by $\beta_{\lambda_{\rm mass}}=\frac{{|\lambda_{\rm mass}}|-|\lambda_{\rm isco}|}{|\lambda_{\rm ibco}|-|\lambda_{\rm isco}|}$ for a fixed $\eta_m$, which is the same value of $\eta_m$ at B in Fig. \ref{homoclinic_region_1} allowing $\lambda_m$ to vary.
Since $ |\lambda_{\rm isco}| \le |\lambda_{\rm mass}| \le |\lambda_{\rm ibco}|$,
$0\le \beta_{\lambda_{\rm mass}} \le 1$.
It is evident that the finite charge of the black hole will enhance the phase difference for both the direct motion ($\lambda_{\rm mass} >0$) and retrograde motion ($\lambda_{\rm mass} < 0$).
Also, for a fixed $Q$, $\Delta\phi^{\rm diff}$
increases with the spin parameter $a$ of the black hole.
This means that the homoclinic orbit moving toward the unstable orbit  can synchronize with the circular motion in terms of the Mino time but in the end with a phase difference.
This may provide useful information to construct the full spectrum of the gravitation waves in the case of EMRIs and is one of the interesting applications from the analytical solutions we obtain in this paper.

\section{equatorial homoclinic orbits}\label{secV}

This section is devoted to the equatorial homoclinic solution, exemplified by the case  C in Fig. \ref{homoclinic_region_1} of the parameter space $\lambda_m$ and $\eta_m$.
The main purpose here is to reduce our exact formulas for nonequatorial orbits to the expressions obtained in the context of the equatorial case.
We will show some important steps to find their consistency.
Since the results of \cite{Levin_2009} are written as a function $r$, we begin with a useful relation between Mino time $\tau_m$ and $r$ given by the inverse of (\ref{r_tau_m_h}),
\begin{equation}\label{tau_m_r}
    \tau_m(r)=\frac{2}{\sqrt{(1-\gamma_m^2)(r_{\rm mss}-r_{m1})(r_{m4}-r_{\rm mss})}}\tanh^{-1}\left({x^H(r)}\right) \, .
\end{equation}
For the proper time ${\sigma_{m}(r)}$, following (\ref{r_eq_particle}) and (\ref{tau'}) with $\Sigma=r^2$ at the equatorial plane we then have
{\begin{equation}
    \sigma_m\left(r\right)-\sigma_{mi}=\int_{r_i}^r\frac{r^2}{\pm_r\sqrt{R_m(r)}}dr=
    \frac{1}{\sqrt{1-\gamma_{m}^2}}I_{2}^{H}\left(\tau_{m}\left(r\right)\right) \,.
\end{equation}}

 The  term of $I^H_2$ given in (\ref{I_2_H}) can be expressed by $r$ using (\ref{tau_m_r}).
 Rearranging them leads to
{\begin{align}
\sigma_m\left(r\right)&-\sigma_{mi}=
\sqrt{\frac{(r-r_{\rm mss})^2(r_{m4}-r)}{(1-\gamma_m^2)(r-r_{m1})}}
+\sqrt{\frac{(r_{\rm mss}-r_{m1})(r_{m4}-r_{\rm mss})}{1-\gamma_m^2}}x^H (r) \notag \\
&+\frac{(r_{m4}-r_{m1})(r_{m1}+2r_{\rm mss}+r_{m4})}{\sqrt{(1-\gamma^2_m)(r_{\rm mss}-r_{m1})(r_{m4}-r_{\rm mss})}}\frac{\sqrt{h}}{h-1}\tanh^{-1}\left(\sqrt{h}\;x^H(r)\right)\notag \\
&-\frac{r_{\rm mss}(r_{m4}-r_{m1})-r_{m1}(r_{m4}+r_{m1})+(r_{m4}-r_{m1})(r_{m1}+2r_{\rm mss}+r_{m4})/(h-1)}{\sqrt{1-\gamma_m^2(r_{\rm mss}-r_{m1})(r_{m4}-r_{\rm mss})}} \tanh^{-1}\left(x^H(r)\right)
%
\end{align}}
Plugging in {$x^H(r)$} and $h$ from (\ref{Upsilon_m_h}) and (\ref{h_pm_h}), and doing a little algebra we arrive at
{\begin{align}\label{sigma_m_r}
\sigma_m\left(r\right)-\sigma_{mi}
&=\sqrt{\frac{\left(r-r_{m1}\right)\left(r_{m4}-r\right)}{1-\gamma_{m}^2}}+\frac{2M}{\left(1-\gamma_{m}^2\right)^{3/2}}\tan^{-1}\sqrt{\frac{r_{m4}-r}{r-r_{m1}}}\notag\\
&+2\sqrt{\frac{\left(r_{\rm mss}-r_{m1}\right)\left(r_{\rm mss}+r_{m1}\right)^2}{\left(1-\gamma_{m}^2\right)\left(r_{m4}-r_{\rm mss}\right)}}\tanh^{-1}\sqrt{\frac{\left(r_{\rm mss}-r_{m1}\right)\left(r_{m4}-r\right)}{\left(r_{m4}-r_{\rm mss}\right)\left(r-r_{m1}\right)}} \, .
\end{align}}
This is the Kerr-Newman's counterpart of equation (B66) in Ref. \cite{Levin_2009}.

Taking the similar steps allows to obtain the equatorial homoclinic orbit in $\phi(r)$ coordinate.
With the help of (\ref{phi_eq_particle}) evaluated at the equatorial plane and (\ref{tau'}), we  first rewrite
\begin{equation}
{\phi}(r)-{\phi_i}={I_\phi}\left(\tau_{m}\left(r\right)\right)+\lambda_{m}\tau_{m}\left(r\right),
\end{equation}
where ${I_\phi}$ is given by (\ref{I_phi_tau_m_h}) and $\tau_m$ by (\ref{tau_m_r}). All the integrals for calculating ${\phi}(r)$ are available from (\ref{I_pm_H}). The result can be written as follows
\begin{align}
    \phi_m(r)-\phi_{mi}=\frac{2}{r_+-r_-}\frac{{\mathcal{A}\tanh^{-1}\left(x^H(r)\right)+\mathcal{B_+}\tanh^{-1}\left(\sqrt{h_+}x^H(r)\right)+\mathcal{B_-}\tanh^{-1}\left(\sqrt{h_-}r^H(r)\right)}}{\sqrt{(1-\gamma^2_m)(r_{{\rm msc}}-r_{m1})(r_{m4}-r_{{\rm msc}})}}\,,
\end{align}
where the coefficients $\mathcal{A}$ and $\mathcal{B_\pm}$ are $r$ independent quantities involving the black hole parameters and the constants of motion of the particle, reading as
%
\begin{figure}[h]
 \centering
 \includegraphics[scale=0.4]{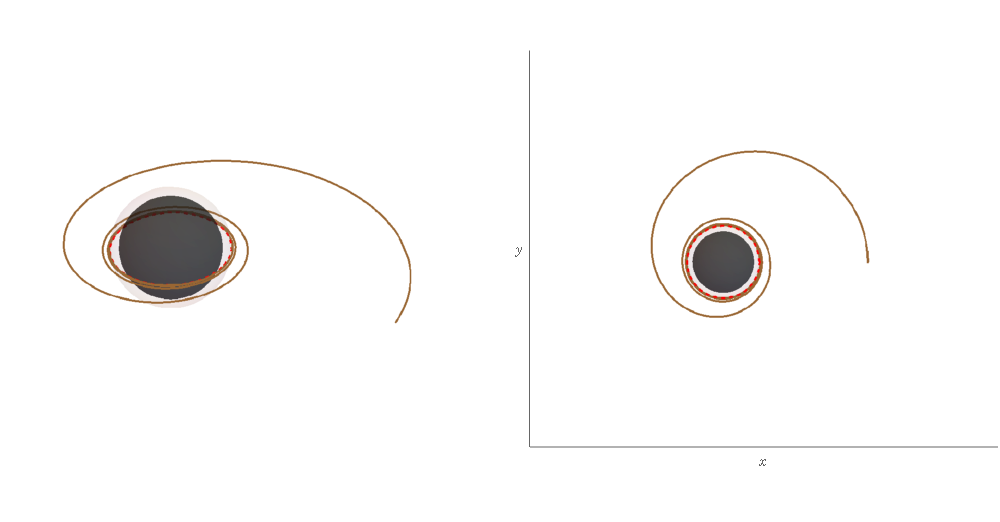}
 \caption{
Illustration of the equatorial homoclinic orbit of case C in Fig. \ref{homoclinic_region_1}.  
{  The red dotted circle is on the equatorial plane.}
 \label{homoclinic_3d}}
 \end{figure}

{\begin{align}
    \mathcal{A}&=\lambda_m\left(r_+-r_-\right)+\frac{2 M a\gamma_m r_+-a^2\lambda_m-a\gamma_m Q^2}{r_{m1}-r_+}\left[ 1 - \frac{r_{m1}-r_{m4}}{(r_{m4}-r_+)(h_+-1)}\right]  \notag \\
    &\quad\quad -\frac{2 M a\gamma_m r_--a^2\lambda_m-a\gamma_m Q^2}{r_{m1}-r_-}\left[ 1 - \frac{r_{m1}-r_{m4}}{(r_{m4}-r_-)(h_--1)}\right] \;,
\end{align}}
and
\begin{align}
    \mathcal{B_\pm}=\pm\frac{(2{M}a\gamma_m r_\pm-a^2{\lambda_m}-a{\gamma_m} Q^2 )(r_{m1}-r_{m4})}{(r_{m1}-r_\pm)(r_{m4}-r_\pm)}\frac{\sqrt{h_\pm}}{h_\pm-1}\;.
\end{align}
%
%
Replacing {$x^H$} and $h_\pm$ gives
\begin{align}
&\;\;\phi_m\left(r\right)-\phi_{mi}=
\notag\\
&2\sqrt{\frac{\left(r_{{\rm msc}}-r_{m1}\right)^3}{\left(1-\gamma_{m}^2\right)\left(r_{m4}-r_{{\rm msc}}\right)}}\frac{r_{{\rm msc}}^2\lambda_{m}+\left(2M r_{{\rm msc}}-Q^2\right)\left(a\gamma_{m}-\lambda_{m}\right)}{\left(r_{{\rm msc}}-r_{+}\right)\left(r_{{\rm msc}}-r_{-}\right)\left(r_{{\rm msc}}-r_{m1}\right)^2}
\tanh^{-1}\sqrt{\frac{\left(r_{{\rm msc}}-r_{m1}\right)\left(r_{m4}-r\right)}{\left(r_{m4}-r_{{\rm msc}}\right)\left(r-r_{m1}\right)}}\notag\\
&-\frac{2}{r_{+}-r_{-}}\frac{\left(2M a\gamma_{m}-r_{-}\lambda_{m}\right)r_{+}-Q^2\left(a\gamma_{m}-\lambda_{m}\right)}{\left(r_{{\rm msc}}-r_{+}\right)\sqrt{\left(1-\gamma_{m}^2\right)\left(r_{+}-r_{m1}\right)\left(r_{m4}-r_{+}\right)}}\tanh^{-1}\sqrt{\frac{\left(r_{+}-r_{m1}\right)\left(r_{m4}-r\right)}{\left(r_{m4}-r_{+}\right)\left(r-r_{m1}\right)}}\notag\\
&+\frac{2}{r_{+}-r_{-}}\frac{\left(2M a\gamma_{m}-r_{+}\lambda_{m}\right)r_{-}-Q^2\left(a\gamma_{m}-\lambda_{m}\right)}{\left(r_{{\rm msc}}-r_{-}\right)\sqrt{\left(1-\gamma_{m}^2\right)\left(r_{-}-r_{m1}\right)\left(r_{m4}-r_{-}\right)}}\tanh^{-1}\sqrt{\frac{\left(r_{-}-r_{m1}\right)\left(r_{m4}-r\right)}{\left(r_{m4}-r_{-}\right)\left(r-r_{m1}\right)}}\label{phim} \, .
\end{align}

The derivation of $t(r)$ is straightforward. From Eq. (\ref{t_eq_particle}) at the equatorial plane and (\ref{tau'}) we have found
\begin{align}
&t_m\left(r\right)-t_{mi}=I_{mt}\left(\tau_{m}\left(r\right)\right)\notag\\
&=\gamma_{m}\sqrt{\frac{\left(r-r_{m1}\right)\left(r_{m4}-r\right)}{1-\gamma_{m}^2}}+\gamma_{m}\frac{r_{m1}+r_{m4}+2\left(r_{{\rm msc}}+2M\right)}{\sqrt{1-\gamma_{m}^2}}\tan^{-1}\sqrt{\frac{r_{m4}-r}{r-r_{m1}}}\notag\\
&+\sqrt{\frac{4\left(r_{{\rm msc}}-r_{m1}\right)^3}{\left(1-\gamma_{m}^2\right)\left(r_{m4}-r_{{\rm msc}}\right)}}\frac{r_{{\rm msc}}^2\left(r_{{\rm msc}}^2+a^2\right)\gamma_{m}+\left(2M r_{{\rm msc}}-Q^2\right)\left(a^2\gamma_{m}-a\lambda_{m}\right)}{\left(r_{{\rm msc}}-r_{+}\right)\left(r_{{\rm msc}}-r_{-}\right)\left(r_{{\rm msc}}-r_{m1}\right)^2} \notag \\
&\quad\quad\quad\quad\times\tanh^{-1}\sqrt{\frac{(r_{{\rm msc}}-r_{m1})\left(r_{m4}-r\right)}{\left(r_{m4}-r_{{\rm msc}}\right)\left(r-r_{m1}\right)}}\notag\\
&-\frac{2\left(2M r_{+}-Q^2\right)}{r_{+}-r_{-}}\frac{2M\gamma_{m}r_{+}-\left(a\lambda_{m}+Q^2\gamma_{m}\right)}{\left(r_{{\rm msc}}-r_{+}\right)\sqrt{\left(1-\gamma_{m}^2\right)\left(r_{+}-r_{m1}\right)\left(r_{m4}-r_{+}\right)}}\tanh^{-1}\sqrt{\frac{\left(r_{+}-r_{m1}\right)\left(r_{m4}-r\right)}{\left(r_{m4}-r_{+}\right)\left(r-r_{m1}\right)}}\notag\\
&+\frac{2\left(2M r_{-}-Q^2\right)}{r_{+}-r_{-}}\frac{2M\gamma_{m}r_{-}-\left(a\lambda_{m}+Q^2\gamma_{m}\right)}{\left(r_{{\rm msc}}-r_{-}\right)\sqrt{\left(1-\gamma_{m}^2\right)\left(r_{-}-r_{m1}\right)\left(r_{m4}-r_{-}\right)}}\tanh^{-1}\sqrt{\frac{\left(r_{-}-r_{m1}\right)\left(r_{m4}-r\right)}{\left(r_{m4}-r_{-}\right)\left(r-r_{m1}\right)}}\, .\label{tm}
\end{align}

%
In summary, the obtained  formulas (\ref{sigma_m_r}), (\ref{phim}), and (\ref{tm}) describe the equatorial homoclinic motion when the parameters and initial conditions are given.
They are consistent with the results given in Appendix B of Ref. \cite{Levin_2009}, as $Q \rightarrow 0$ that results in $r_{m1}=0$.
Figure \ref{homoclinic_3d} shows an example of such an orbit for the solution given above. The parameters of the plot correspond to case C in Fig. \ref{homoclinic_region_1}.

 \begin{figure}[h]
 \centering
 \includegraphics[width=0.99\columnwidth=0.99,trim=20 400 20 400,clip]{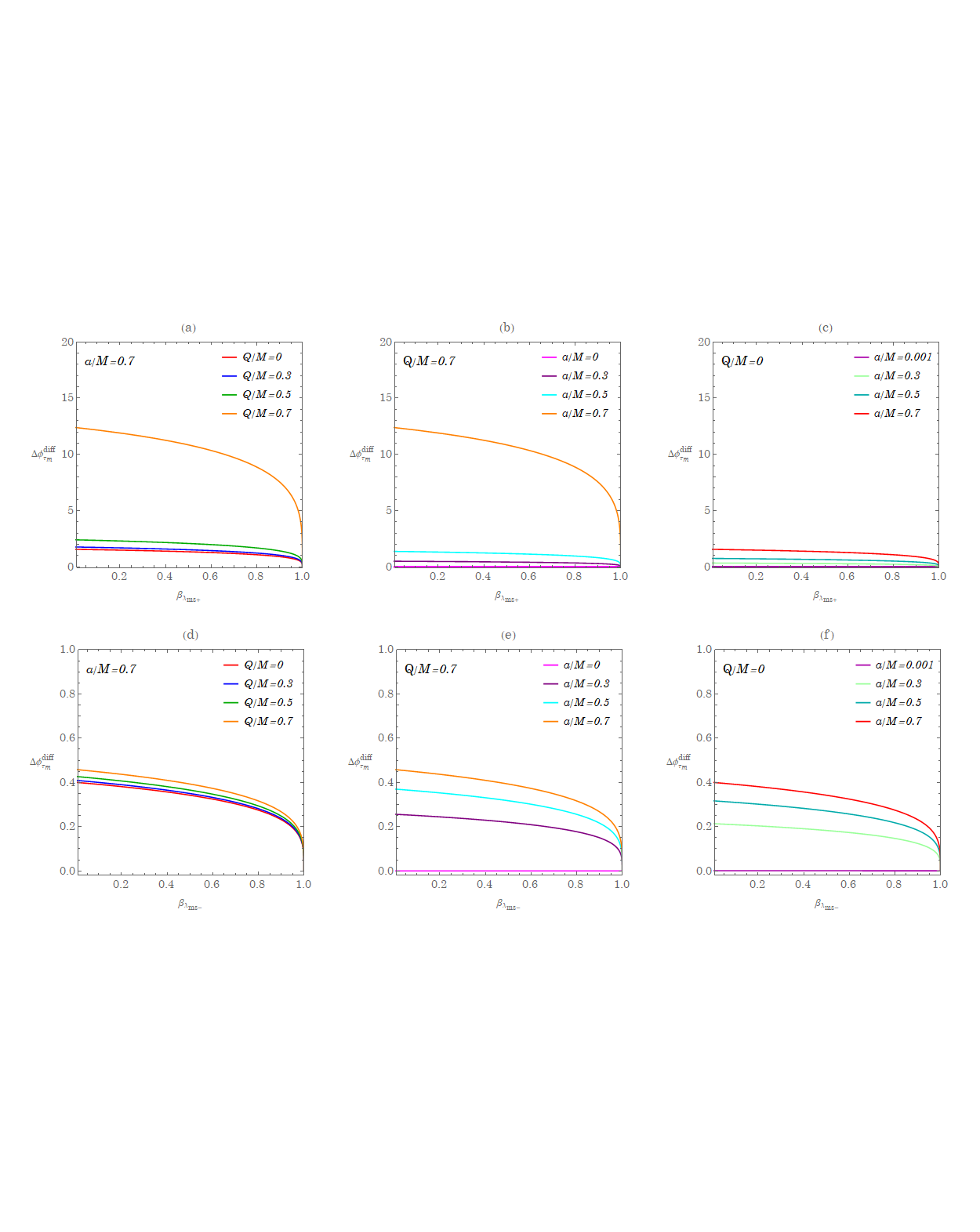}
 \caption{
  {Phase difference $\Delta \phi_{\tau_{m}}^{\rm diff}=\phi^{\rm circ}-\phi_m$ as a function of $\beta_{\lambda_{\rm msc}}=\frac{|\lambda_{\rm ibco}|-|\lambda_{\rm msc}|}{|\lambda_{\rm ibco}|-|\lambda_{\rm isco}|}$ in (\ref{phase_diff_tau}) for the case of equatorial solutions.
The upper plots are for the direct motion and the lower plots are for the retrograde motion.}
 \label{Phase_difference_a}}
 \end{figure}
%
Figure \ref{Phase_difference_a} considers the phase difference between the homoclinic orbit and the circular motion on the equatorial plane in terms of the Mino time by taking $\eta_{\rm mass} \rightarrow 0$ limit in Eq.(\ref{phase_diff_tau}).
The results of the Kerr cases are consistent with those in \cite{Levin_2009}. In addition, the study is extended to the black holes with finite charge to see the enhancement to the phase difference.

\section{Summary and outlook}

\nopagebreak

In this paper, we analytically derive the homoclinic solutions for the  general nonequatorial orbits in the Kerr-Newman black holes.
In the limits of $Q\rightarrow 0$ and restricting on the equatorial plane, the obtained solution reduces to the one obtained in \cite{Levin_2009}.
We then  consider the homoclinic orbits with zero azimuthal angular momentum.
The evolution of the $\phi$ angle is solely due to the frame dragging effects.
The larger value of the black hole spin $a$ gives relatively large dragging
effects and moreover, the charge Q of the black hole gives the boost to it.
Also, we compute the phase difference between the { nonequatorial homoclinic solutions  and the unstable spherical motion, giving the finite phase difference in terms of the Mino time as $ \tau_m \rightarrow \infty$ ( $ t\rightarrow \infty$)}, which  might provide an interesting information to construct the full spectrum of the gravitational waves emitted by the EMRIs.

The existence of the homoclinic solutions which is separatrix between the motions of the inspiral types and plunging into the black holes gives the system to become chaotic when a random kick is acted on the particle \cite{Liu}.
The Lyapunov exponent for the chaotic system has been extensively investigated from the viewpoint of
black hole systems with a probe particle based upon the AdS/CFT correspondence.
According to the AdS/CFT duality in the case of the Schwarzschild black hole,  it is claimed  that there exists the bound on the Lyapunov exponent of such a probe particle \cite{Mald}.
Explicit calculations of the Lyapunov exponent will give insight on whether this conjecture of the bound on the Lyapunov exponent can be applied to the Kerr-Newman black holes.
As long as any one of the evaluating Lyapunov exponents is a nonzero positive value at large
time, we could probably see it as chaotic dynamics since the initial infinitesimal difference
experiences exponential divergence eventually.
Lastly, we can consider the system to experience an external time-dependent random noise from the random kicks, and are able to construct a Melnikov function.
By taking advantage of the analytical expression of the homoclinic solution, the Melnikov function could be analytically solved. 
The simple zeros of the Melnikov function  imply the occurrence of chaotic dynamics \cite{Bom,Syu}.

\nopagebreak
\appendix
\nopagebreak
\section{Angular potential $\Theta(\theta)$ and the integrals ${G_\theta}$, ${G_\phi}$, and ${G_t}$}
\nopagebreak
%
For the sake of completeness, we summarize the relevant results related to the $\Theta_m$ potential in the $\theta$ direction \cite{Gralla_2020a,Wang_2022}.
The angular potential (\ref{ThetaPotential}) for the particle can be rewritten in terms of $u=\cos^2 \theta$ as
\begin{align}
(1-u)\Theta_m(u)=-{a}^2\left(\gamma_m^2-1\right)u^2+\left[{a}^2\left(\gamma_m^2-1\right)-\left(\eta_m+\lambda_m^2\right)\right]u+\eta_m \, .
\end{align}
The positivity of the $\Theta_m$ potential required $\lambda_m$ and $\eta_M$ restricted in some parameter space (see Fig. 9 in \cite{Wang_2022}).
The boundaries are determined from the roots of $\Theta_m(\theta)=0$,
\begin{align}\label{u_m}
u_{{m\pm}}=\frac{\Delta_{{m\theta}}\pm\sqrt{\Delta_{{m\theta}}^2+4\,{a}^2\, \eta_m}}{2{a}^2}\,,
\quad
\Delta_{m \theta}={a}^2-\frac{\eta_m+\lambda_m^2}{\gamma_m^2-1}\, .
\end{align}

We consider the non-negative $\eta_m \ge 0$ for the orbits of the black hole exteriors and that requires the full trajectories lying in the Kerr-Newman black hole exterior \cite{Wang_2022}.
For $\eta_m >0$ and nonzero $\lambda_m$ of the bound motion of the homoclinic orbits, $1>u_{{m+}}>0$  is the only positive root that in turn gives two roots at $\theta_{m+}=\cos^{-1}\left(-\sqrt{u_{{m+}}}\right), \theta_{m-}=\cos^{-1}\left(\sqrt{u_{{m+}}}\right)$.
The particle travels between the southern and northern hemispheres crossing the equator.
The corresponding solutions are expressed here as
\begin{align}
\tau_m={G_\theta}=p ({\mathcal{G}_{\theta_{m+}}}- {\mathcal{G}_{\theta_{m-}}}) + \nu_{\theta_i} \left[(-1)^p{\mathcal{G}_{\theta}}-{\mathcal{G}_{\theta_i}}\right], \label{tau_G_theta_m_a}
\end{align}
where the trajectory passes through the turning point $p$ times and $\nu_{\theta_i}={\rm sign}\left(\frac{d\theta_{i}}{d{\tau_m}}\right)$.
The function ${\mathcal{G}_{\theta}}$ is
\be \label{g_theta_m_a}
{\mathcal{G}_{\theta}}=-\frac{1}{\sqrt{-u_{m-}{a}^2\left(\gamma_m^2-1\right)}}F\left(\sin^{-1}\left(\frac{\cos\theta}{\sqrt{u_{m+}}}\right) \left|\frac{u_{m+}}{u_{m-}}\right)\right.\, .
\ee
Inversion of (\ref{r_theta}) gives $\theta(\tau_m)$ as \cite{Gralla_2020a,Wang_2022}
\be \label{theta_tau_m_a}
\theta(\tau_m)=\cos^{-1}\left(-\nu_{\theta_i}\sqrt{u_{m+}}{\rm sn}\left(\sqrt{-u_{m -}{a}^2\left(\gamma_m^2-1\right)}\left(\tau_m+\nu_{\theta_i}{\mathcal{G}_{\theta_i}}\right)\left|\frac{u_{m+}}{u_{m-}}\right)\right.\right)
\ee
involving the Jacobi elliptic sine function \cite{Abramowitz}.

The other two integrals related to $G(\theta)$ that contribute to the solutions of $\phi$ and $t$ in (\ref{phi}) and (\ref{t}) are
\begin{align}
{G_\phi}(\tau_m)&=\frac{1}{\sqrt{-u_{m -}{a}^2\left(\gamma_m^2-1\right)}}\Pi\left(u_{m+};{\rm am}\left(\sqrt{-u_{m-}{a}^2\left(\gamma_m^2-1\right)}\left(\tau_m+\nu_{\theta_i}\mathcal{G}_{\theta_i}\right)\left|\frac{u_{m+}}{u_{m-}}\right)\right.\left|\frac{u_{m+}}{u_{m-}}\right)\right. \notag \\ &\quad-\nu_{\theta_i}{\mathcal{G}_{\phi_i}} \,,\label{G_phi_tau_m_a}
\end{align}
where
\begin{equation}
    \mathcal{G}_{\phi_i}=-\frac{1}{\sqrt{-u_{m-}{a}^2\left(\gamma_m^2-1\right)}}\Pi\left(u_{m+};\sin^{-1}\left(\frac{\cos\theta_i}{\sqrt{u_{m+}}}\right)\left|\frac{u_{m+}}{u_{m-}}\right)\right.
\end{equation}
and
\begin{align}
   {G_t}(\tau_m)&=-\frac{2u_{m+}}{\sqrt{-u_{m-}{a}^2\left(\gamma_m^2-1\right)}}E'\left({\rm am}\left(\sqrt{-u_{m-}{a}^2\left(\gamma_m^2-1\right)}\left(\tau_m+\nu_{\theta_i}{\mathcal{G}_{\theta_i}}\right)\left|\frac{u_{m+}}{u_{m-}}\right)\right.\left|\frac{u_{m+}}{u_{m-}}\right)\right. \notag \\
   &\quad -\nu_{\theta_i}\mathcal{G}_{t_i}\,,\label{G_t_tau_m_a}
\end{align}
where
\begin{equation}
   \mathcal{G}_{t_i}=\frac{2u_{m+}}{\sqrt{-u_{m-}{a}^2\left(\gamma_m^2-1\right)}}E'\left(\sin^{-1}\left(\frac
{\cos\theta_i}{\sqrt{u_{+}}}\right)\left|{\frac{u_{m+}}{u_{m-}}}\right)\right.\,.
\end{equation}
The results above also involve incomplete elliptic integrals of the second  and third kinds. The derivative of $E(\varphi\mid k)$ can be calculated through
\begin{align}
E'\left(\varphi\left|k\right)\right.=\partial_k E\left(\varphi\left|k\right)\right.=\frac{E\left(\varphi\left|k\right)\right.-F\left(\varphi\left|k\right)\right.}{2 k} \,.
\end{align}
 Notice that since  $0< \frac{u_{m+}}{u_{m-}}<1$ for the bound motion, the involved elliptic functions are all real valued and finite. In particular, for $\eta_m=0$, the orbits are on the equatorial plane at $\theta=\frac{\pi}{2}$.

\section{Radial potential $R_m(r)$ and the integrals ${I_\theta}$, ${I_\phi}$, and ${I_t}$}\label{app_B}

As for the radial potential, we examine the parameter space to have positive $R_m({r})$. It is a quartic function of the form
\cite{Wang_2022}
\begin{align}\label{R_m}
R_m({r})=S_m {r}^4+T_m {r}^3+U_m {r}^2+V_m {r}+W_m\,,
\end{align}
where the coefficients are written in terms of the black hole parameters and the constants of motion,
{\begin{align}
&S_m=\gamma_m^2-1,\\
&T_m=2M,\\
&U_m={a}^2\left(\gamma_m^2-1\right)-{Q}^2-\eta_m-\lambda_m^2,\\
&V_m=2M\Bigl[\left({a}\gamma_m-\lambda_m\right)^2+\eta_m\Bigr],\\
&W_m=- {a}^2\eta_m-{Q}^2\Bigl[\left({a}\gamma_m-\lambda_m\right)^2+\eta_m\Bigr]\,.
\end{align}}
We name the roots as $R_m({r})=\left(\gamma_m^2-1\right)({r}-r_{m 1})({r}-r_{m 2}) ({r}-r_{m 3}) ({r}-r_{m 4})$, being $\left(r_{m 4}>r_{m 3}>r_{m 2}>r_{m 1}\right)$.
The explicit formulas of the roots are cumbersome, we summarize them below:
\be
r_{m 1}=-\frac{M}{2\left(\gamma_m^2-1\right)}-z_m-\sqrt{-\hspace*{1mm}\frac{{X}_m}{2}-
z_m^2+\frac{{Y}_m}{4z_m}}\,,
\ee
\be
r_{m 2}=-\frac{M}{2\left(\gamma_m^2-1\right)}-z_m+\sqrt{-\hspace*{1mm}\frac{{X}_m}{2}-z_m^2+\frac{{Y}_m}{4z_m}}\,,
\ee
\be
r_{m 3}=-\frac{M}{2\left(\gamma_m^2-1\right)}+z_m-\sqrt{-\hspace*{1mm}\frac{{X}_m}{2}-z_m^2-\frac{{Y}_m}{4z_m}}\,,
\ee
\be
r_{m 4}=-\frac{M}{2\left(\gamma_m^2-1\right)}+z_m+\sqrt{-\hspace*{1mm}\frac{{X}_m}{2}-z_m^2-\frac{{Y}_m}{4z_m}}\,,
\ee
where
\be
z_m=\sqrt{\frac{\Omega_{m +}+\Omega_{m -}-\frac{{X}_m}{3}}{2}}\; ,
\quad
\Omega_{m \pm}=\sqrt[3]{-\hspace*{1mm}\frac{{\varkappa}_m}{2}\pm\sqrt{\left(\frac{{\varpi}_m}{3}\right)^3+\left(\frac{{\varkappa}_m}{2}\right)^2}}\,
\ee
with
\be
{\varpi}_m=-\hspace*{1mm}\frac{{X}_m^2}{12}-{Z}_m \, , \quad\quad
{\varkappa}_m=-\hspace*{1mm}\frac{{X}_m}{3}\left[\left(\frac{{X}_m}{6}\right)^2-{Z}_m\right]-\hspace*{1mm}\frac{{Y}_m^2}{8}\,.
\ee
$X_m$, $Y_m$, and $Z_m$ are the short notation for
{\begin{align}
&{X}_m=\frac{8U_m S_m -3T_m^2}{8S_m^2}\,,\\
&{Y}_m=\frac{T_m^3-4U_m T_m S_m+8V_m S_m^2}{8S_m^3}\,,\\
&{Z}_m=\frac{-3T_m^4+256W_m S_m^3-64V_m T_m S_m^2+16U_m T_m^2S_m}{256S_m^4}\,.
\end{align}}

{The sum of the roots satisfies the relation $r_{m 1}+r_{m 2}+r_{m 3}+r_{m 4}=-\frac{2}{\gamma_m^2-1}\;$.} Some of the feature of $R_m$ is illustrated in Fig. \ref{phase_portrait}. See also Fig. 10 in \cite{Wang_2022}.

We at this stage show an analytical solution of bound orbits, where $r_{m3}\leq r_i\leq r_{m4}$, which
is suited to constructing the homoclinic solution discussed in Sec. \ref{sec_IIC}.
The bound solution of $r(\tau_m)$ is given by the inversion of (\ref{t}),
\be
r(\tau_m)=\frac{r_{m4}(r_{m1}-r_{m3})-r_{m1}(r_{m4}-r_{m3}){\rm sn}^2\left(X^{B}(\tau_m)\left|k^{B}\right)\right.}{(r_{m1}-r_{m3})-(r_{m4}-r_{m3}){\rm sn}^2\left(X^{B}(\tau_m)\left|k^{B}\right)\right.}\label{r_tau_m_B}\,,
\ee
where
\be
X^{B}(\tau_m)=\frac{\sqrt{\left(1-\gamma_m^2\right)(r_{m3}-r_{m1})(r_{m4}-r_{m2})}}{2}\tau_m-\nu_{r_i} F\Bigg(\sin^{-1}\left(\sqrt{\frac{(r_{i}-r_{m4})(r_{m1}-r_{m3})}{(r_{i}-r_{m1})(r_{m4}-r_{m3})}}\right)\left|k^{B}\Bigg)\right.\,\label{X_B_tau_m}\\
\ee
and
\be
k^{B}=\frac{(r_{m4}-r_{m3})(r_{m2}-r_{m1})}{(r_{m4}-r_{m2})(r_{m3}-r_{m1})} \, .
\ee
with $\nu_{r_i}={\rm sign}\left(\frac{dr_{i}}{d{\tau_m}}\right)$ and $\rm sn$ denoting the Jacobi elliptic sine function.

The other integrals involving the radial potential $R_m(r)$ are obtained as
\be
I^B_{\phi}(\tau_m)=\frac{\gamma_m}{\sqrt{1-\gamma_m^2}}\frac{2Ma}{r_{+}-r_{-}}\left[\left(r_{+}-\frac{a\left(\frac{\lambda_m}{\gamma_m}\right)+Q^2}{2M}\right)I_{+}^{B}(\tau_m)-\left(r_{-}-\frac{a\left(\frac{\lambda_m}{\gamma_m}\right)+Q^2}{2M}\right)I_{-}^{B}(\tau_m)\right]\label{I_phi_tau_m_b}\,,
\ee
\begin{align}
&I^B_{t}(\tau_m)=\frac{\gamma_m}{\sqrt{1-\gamma_m^2}}\left\lbrace\frac{4M^2}{r_{+}-r_{-}}\left[\left(r_{+}-\frac{Q^2}{2M}\right)\left(r_{+}-\frac{a\left(\frac{\lambda_m}{\gamma_m}\right)+Q^2}{2M}\right)I_{+}^{B}(\tau_m)\right.\right.\notag\\
&\quad\quad\left.\left.-\left(r_{-}-\frac{Q^2}{2M}\right)\left(r_{-}-\frac{a\left(\frac{\lambda_m}{\gamma_m}\right)+Q^2}{2M}\right)I_{-}^{B}(\tau_m)\right]+2MI_{1}^{B}(\tau_m)+I_{2}^{B}(\tau_m) \label{I_t_tau_m_b}\right\rbrace+\left(4{M^2}-Q^2\right)\gamma_{m}\tau_m\,,
\end{align}
where
{\begin{align}
&I_{\pm}^{B}(\tau_m)=\frac{2}{\sqrt{(r_{m3}-r_{m1})(r_{m4}-r_{m2})}}\left[\frac{X^{B}(\tau_m)}{r_{m1}-r_{\pm}}+\frac{r_{m1}-r_{m4}}{(r_{m1}-r_{\pm})(r_{m4}-r_{\pm})}\Pi\left(\beta_{\pm};\Upsilon_{\tau_m}^{B}\left|k^{B}\right)\right.\right]+\nu_{r_i}\mathcal{I}_{\pm_i}^{B}\\
&\mathcal{I}_{\pm_i}^{B}=\frac{2}{\sqrt{(r_{m3}-r_{m1})(r_{m4}-r_{m2})}}\left[\frac{F\left(\Upsilon_{r_i}^{B}\mid k_{B}\right)}{r_{m1}-r_{\pm}}+\frac{r_{m1}-r_{m4}}{(r_{m1}-r_{\pm})(r_{m4}-r_{\pm})}\Pi\left(\beta_{\pm};\Upsilon_{r_i}^{B}\left|k_{B}\right)\right.\right]\\
&I_{1}^{B}(\tau_m)=\frac{2}{\sqrt{(r_{m3}-r_{m1})(r_{m4}-r_{m2})}}\left[r_{m1}X^{B}(\tau_m)+(r_{m4}-r_{m1})\Pi\left(\beta;\Upsilon_{\tau_m}^{B}\left|k^{B}\right)\right.\right]+\nu_{r_i}\mathcal{I}_{1_i}^{B}\\
&\mathcal{I}_{1_i}^{B}=\frac{2}{\sqrt{(r_{m3}-r_{m1})(r_{m4}-r_{m2})}}\left[r_{m1}F\left(\Upsilon_{r_i}^{B}\left|k_{B}\right)\right.+(r_{m4}-r_{m1})\Pi\left(\beta;\Upsilon_{r_i}^{B}\left|k_{B}\right)\right.\right]\\
&I_{2}^{B}(\tau_m)=-\nu_{r}\frac{\sqrt{\left(r(\tau_m)-r_{m1}\right)\left(r(\tau_m)-r_{m2}\right)\left(r(\tau_m)-r_{m3}\right)\left(r_{m4}-r(\tau_m)\right)}}{r(\tau_m)-r_{m1}}\notag\\
&\quad\quad-\frac{r_{m3}\left(r_{m4}-r_{m1}\right)-r_{m1}\left(r_{m4}+r_{m1}\right)}{\sqrt{(r_{m3}-r_{m1})(r_{m4}-r_{m2})}}X^{B}(\tau_m)+\sqrt{(r_{m3}-r_{m1})(r_{m4}-r_{m2})}E\left(\Upsilon_{\tau_m}^{B}\left|k^{B}\right)\right.\notag\\
&\quad\quad+\frac{\left(r_{m4}-r_{m1}\right)\left(r_{m1}+r_{m2}+r_{m3}+r_{m4}\right)}{\sqrt{(r_{m3}-r_{m1})(r_{m4}-r_{m2})}}\Pi\left(\beta;\Upsilon_{\tau_m}^{B}\left|k^{B}\right)\right.+\nu_{r_i}\mathcal{I}_{2_i}^{B}
\end{align}}
\begin{align}
&\mathcal{I}_{2_i}^{B}=\frac{\sqrt{\left(r_i-r_{m1}\right)\left(r_i-r_{m2}\right)\left(r_i-r_{m3}\right)\left(r_{m4}-r_i\right)}}{r_i-r_{m1}}-\frac{r_{m3}\left(r_{m4}-r_{m1}\right)-r_{m1}\left(r_{m4}+r_{m1}\right)}{\sqrt{(r_{m3}-r_{m1})(r_{m4}-r_{m2})}}F\left(\Upsilon_{r_i}^{B}\left|k^{B}\right)\right.\notag\\
&\quad\quad+\sqrt{(r_{m3}-r_{m1})(r_{m4}-r_{m2})}E\left(\Upsilon_{r_i}^{B}\left|k^{B}\right)\right.+\frac{\left(r_{m4}-r_{m1}\right)\left(r_{m1}+r_{m2}+r_{m3}+r_{m4}\right)}{\sqrt{(r_{m3}-r_{m1})(r_{m4}-r_{m2})}}\Pi\left(\beta;\Upsilon_{r_i}^{B}\left|k^{B}\right)\right.
\end{align}

The parameters in the equations above are
\begin{align}\label{Upsilon_m_b}
&\Upsilon_{r}^{B}=\sin^{-1}\sqrt{\frac{(r-r_{m4})(r_{m1}-r_{m3})}{(r-r_{m1})(r_{m4}-r_{m3})}},\hspace*{4mm}\Upsilon_{\tau_m}^{B}={\rm am}\bigl(X^{B}(\tau_m)\left|k_{B}\bigr)\right.\\
&\nu_{r}={\rm sign}\left(\frac{dr(\tau_m)}{d\tau_m}\right),\hspace*{4mm}\beta^B_{\pm}=\frac{(r_{m1}-r_{\pm})(r_{m4}-r_{m3})}{(r_{m4}-r_{\pm})(r_{m1}-r_{m3})},\hspace*{4mm}\beta^B=\frac{r_{m4}-r_{m3}}{r_{m1}-r_{m3}}\, .
\end{align}

\begin{acknowledgments}
This work was supported in part by the National Science and Technology Council (NSTC) of Taiwan, Republic of China.
\end{acknowledgments}

\end{document}